\documentclass[
  reprint,
  superscriptaddress,
  amsmath,
  amssymb,
  aps,
  prl,
  longbibliography,
]{revtex4-2}

\usepackage[T1]{fontenc}
\usepackage[utf8]{inputenc}
\usepackage{bm}
\usepackage{graphicx}
\usepackage{hyperref}

\usepackage{xcolor}
\usepackage{ulem}

\newcommand{\dd}{\mathrm{d}}
\newcommand{\tr}{\mathrm{Tr}}
\newcommand{\re}{\mathrm{Re}}
\newcommand{\im}{\mathrm{Im}}

\begin{document}

\title{Boundary Completion of Vacuum Persistence Probability}

\author{Yu Zhou}
\email{YuZhou\_@buaa.edu.cn}
\affiliation{Center for Gravitational Physics, Department of Space Science,\\ Beihang University, Beijing 100191, China}

\author{Hai-Qing Zhang}
\email{hqzhang@buaa.edu.cn}
\affiliation{Center for Gravitational Physics, Department of Space Science,\\ Beihang University, Beijing 100191, China}
\affiliation{Peng Huanwu Collaborative Center for Research and Education,\\ Beihang University, Beijing
100191, China}

\date{\today}

\begin{abstract}
The vacuum persistence probability, encoded in the imaginary part of the in-out effective action, is a basic measure of vacuum instability and particle production. However, its evaluation may appear prescription-dependent: the Bogoliubov prescription and the Green's function prescription can yield different results. This discrepancy reflects the incomplete treatment of the endpoint vacuum states in the conventional Green's function reconstruction of the effective action. We establish a boundary-completed reconstruction that consistently incorporates the bulk Green's function and the endpoint vacuum contributions, thereby determining the structure of the complete in-out effective action. The resulting vacuum persistence probability is uniquely determined and agrees with the Bogoliubov result.
\end{abstract}

\maketitle

\paragraph{Introduction.}
Particle production from the vacuum is a characteristic prediction of quantum field theory in external backgrounds and curved spacetime, with well-known examples including the Schwinger effect \cite{Schwinger:1951nm}, Hawking radiation \cite{Hawking:1975vcx}, and cosmological particle production \cite{Parker:1968mv,Parker:1969}.
Such processes can be characterized by the in-out vacuum persistence amplitude
\begin{align}
Z=\langle 0,{\rm out}|0,{\rm in}\rangle\equiv e^{iW},
\end{align}
which defines the in-out effective action $W=-i\log Z$ \cite{DEWITT1975}. Its imaginary part determines the vacuum persistence probability,
$P_{\rm vac}=|Z|^2=e^{-2\rm{Im}W}$.

Two standard approaches are commonly used to determine the effective action. The Bogoliubov approach expresses it in terms of the transformation between the in- and out-mode bases,
\begin{align}\label{eq:BogoliubovFormula}
W_{\rm B}=\frac{i}{2}\tr\log\alpha ,
\end{align}
up to a real phase \cite{DEWITT1975,DeWitt:2003pm,Nikishov_2003}. The Green's function approach instead reconstructs the effective action from the coincident in-out Feynman Green's function \cite{Parker_Toms_2009,Birrell:1982ix},
\begin{align}\label{eq:GreenFormula}
W_{\rm G}
=-\frac{1}{2}\int_{\Omega}\dd^{d+1}x\sqrt{|g|}
\int_{+\infty}^{m^2}\dd\bar m^2
G_{\rm F}(x,x;\bar m^2),
\end{align}
or equivalently from the heat kernel \cite{Vassilevich:2003xt}. These two constructions are commonly regarded as equivalent representations of the same in-out effective action \cite{Ambjorn:1983ne} and have both been widely used in studies of particle production \cite{Kim:2009aq,Chen:2026trv,Wondrak:2023zdi,zhou2024,Fecit:2025kqb,Dunne:2004nc,Ford_2021,RevModPhys.96.045005}.

However, this presumed equivalence is challenged by several well-known examples. In de Sitter spacetime, the Bogoliubov approach \cite{Mottola1985} and the conventional Green's function approach \cite{Akhmedov:2009ta,Akhmedov_2019,zhou2024} can yield different imaginary parts of the effective action \cite{Akhmedov:2024qvi,Zhou:2025jwm}. Even in the flat space Schwinger effect, their agreement is not automatic but requires an additional identification of the longitudinal momentum integral with the interaction time \cite{Nikishov:1969tt,Anderson2018}. 
These discrepancies raise the question of why the two constructions differ and which one corresponds to the physical vacuum persistence amplitude.

An important step toward resolving this issue was taken in Ref.~\cite{Akhmedov:2024qvi}, which analyzed the discrepancy in expanding spacetimes. It emphasized that the in-out functional integral contains not only the bulk action but also the endpoint wavefunctionals specifying the initial and final vacua, whereas the Green's function reconstruction accounts only for the bulk contribution. Including the endpoint contributions restores agreement with the Bogoliubov result for background fields acting over a finite time interval. For background fields that persist eternally, however, the time and momentum integrals in the bulk contribution become divergent, so the interchange of their order required in the finite-time analysis is no longer generally justified. Moreover, the endpoint contributions may also fail to admit well-defined asymptotic limits. The finite-time agreement therefore does not directly settle the eternal case.

This suggests formulating the comparison within a common regularization, before assigning separate meanings to the individual divergent contributions. In our previous analysis of de Sitter spacetime \cite{Zhou:2025jwm}, we introduced explicit late-time and momentum cutoffs and found that the Green's function reconstruction in fact yields a regulator-dependent family of results. The conventional Green's function and Bogoliubov results emerge from the same regulated expression as two opposite regularization limits. That analysis, however, did not determine a unique physical result, nor did it include the endpoint vacuum wavefunctionals.

These results leave two closely related questions. What does the mass-integrated Green's function \eqref{eq:GreenFormula} actually reconstruct? And to what extent is this restoration of agreement universal once the endpoint vacuum wavefunctionals are consistently included within the same regularization?

In this Letter, we identify a boundary-completion structure of the in-out effective action for scalar fields governed by quadratic actions, showing how the Green's function and endpoint vacuum contributions combine to recover the Bogoliubov vacuum persistence probability. At the level of mass variation, the central structure is that the complete effective action contains not only the Green's function contribution but also that from the initial and final vacuum wavefunctionals,
\begin{align}
\partial_{m^2}W
=
\partial_{m^2}W_{\rm G}
+
\partial_{m^2}W_{\Psi}.
\end{align}
The Green's function result differs from the Bogoliubov result by additional boundary contributions,
\begin{align}
\partial_{m^2}W_{\rm G}
=
\partial_{m^2}W_{\rm B}
-\mathcal B_{\rm in}
-\mathcal B_{\rm out},
\end{align}
while the endpoint vacuum wavefunctionals contribute the same boundary terms with the opposite sign,
\begin{align}
\partial_{m^2}W_{\Psi}
=
+\mathcal B_{\rm in}
+\mathcal B_{\rm out}
+\partial_{m^2}\Theta ,
\qquad \Theta\in\mathbb R.
\end{align}
The boundary contributions therefore cancel in the complete effective action, leaving only an arbitrary real phase in addition to the Bogoliubov result.
Consequently, the vacuum persistence probability is uniquely fixed by the Bogoliubov coefficients. Crucially, this cancellation occurs within the same regularization before the regulator is removed, so the separate contributions need not possess regulator-independent limits.

\paragraph{In-out amplitude with endpoint states.}
We begin with the path-integral representation of the vacuum persistence amplitude, keeping the boundary wavefunctionals explicit \cite{Weinberg:1995mt}:
\begin{align}
e^{iW}=Z=\int_{\Omega} \mathcal{D}\phi \Psi_{\rm out}^*[\phi(\Sigma_{\rm out})]\Psi_{\rm in}[\phi(\Sigma_{\rm in})] e^{i S_m[\phi]}.
\label{eq:complete_amplitude}
\end{align}
Here, we assume that $\Omega$ is globally hyperbolic, with initial and final Cauchy surfaces $\Sigma_{\rm in}$ and $\Sigma_{\rm out}$ as its temporal boundaries.  The wavefunctionals $\Psi_{\rm in}$ and $\Psi_{\rm out}$ are the Schrödinger representations of the in- and out-vacua, evaluated on the restrictions of the integration field to the corresponding Cauchy surfaces.  They specify the temporal boundary conditions of the in-out problem, while possible spatial boundary conditions are assumed to ensure sufficiently fast decay of the fields and hence yield no contributions to the integral.
To balance generality and simplicity, we focus our discussion on the real scalar field of mass $m$, with action
$S_m[\phi]=-\frac{1}{2}
\int_{\Omega}\dd^{d+1}x\sqrt{|g|}
\left[g^{\mu\nu}\nabla_\mu\phi\nabla_\nu\phi+m^2\phi^2\right].$
The same framework also applies, with appropriate modifications, to scalar fields coupled to prescribed classical electromagnetic backgrounds, in which the theory remains quadratic in the dynamical field. A minimal $0+1$-dimensional setting based on a time-dependent harmonic oscillator is presented in the Supplemental Material Sec.~III A \cite{supplemental} as a simpler illustration of the underlying mechanism.

The Green's function prescription reconstructs the effective action through its mass variation.  For the complete amplitude \eqref{eq:complete_amplitude}, this variation has two origins: the explicit mass dependence of the bulk action and the mass dependence of the endpoint vacuum wavefunctionals, i.e., 
$\partial_{m^2}W=\left\langle \partial_{m^2}S_m\right\rangle
-i\left\langle \partial_{m^2}\log\Psi_{\rm in}\right\rangle
-i\left\langle \partial_{m^2}\log\Psi_{\rm out}^{*}\right\rangle$ \cite{Schwinger:1951ex}.
Here, $\langle\cdots\rangle$ denotes the expectation value with respect to the in-out path-integral such that
$\langle\mathcal{O}\rangle=Z^{-1}\int_{\Omega}\mathcal{D}\phi\,\mathcal{O}\,\Psi_{\rm out}^{*}\Psi_{\rm in}e^{iS_m[\phi]}$.
The bulk-action variation is precisely given by the coincident in-out Feynman Green's function \cite{Akhmedov_2019},
\begin{align}
\left\langle \partial_{m^2}S_m\right\rangle
=
-\frac{1}{2}
\int_{\Omega}\dd^{d+1}x\sqrt{|g|}\,
G_{\rm F}(x,x)=\partial_{m^2}W_{\rm G},
\label{eq:green_bulk_variation}
\end{align}
where $G_{\rm F}(x,x')=\langle\phi(x)\phi(x')\rangle$ is the in-out Feynman Green's function in the path-integral representation. Thus, the complete mass variation of the effective action $W$ naturally separates into a bulk part and an endpoint part:
\begin{align}
\partial_{m^2}W=\partial_{m^2}W_{\rm G}+\partial_{m^2}W_{\Psi},
\label{eq:complete_variation}
\end{align}
where $\partial_{m^2}W_{\Psi}\equiv-i\left\langle \partial_{m^2}\log\Psi_{\rm in}\right\rangle
-i\left\langle \partial_{m^2}\log\Psi_{\rm out}^{*}\right\rangle$ collects the endpoint contributions from the vacuum wavefunctionals.
After integration over $m^2$, this formally gives 
$W(m^2)=W_{\rm G}(m^2)+W_{\Psi}(m^2)+W_0$,
where $W_0$ is independent of $m^2$ and is fixed by the choice of reference condition. The Green's function prescription \eqref{eq:GreenFormula} therefore reconstructs only the bulk-action response $W_{\rm G}$.  Whether it equals the full in-out effective action depends on the endpoint contribution $W_\Psi$, which will be evaluated later.

\paragraph{Boundary obstruction in the Green's function prescription.}
We next isolate what the bulk Green's function prescription computes.  The result will be the Bogoliubov mass variation in addition to terms that live entirely on the boundaries.
Let $\{u_k^{\rm in},u_k^{{\rm in}*}\}$ and $\{u_{k'}^{\rm out},u_{k'}^{{\rm out}*}\}$ be two complete bases of solutions to the scalar field equation $ (-\nabla_{\mu}\nabla^{\mu}+m^2)u=0 $. The in-modes $u_k^{\rm in}$ are chosen to be positive-frequency near $\Sigma_{\rm in}$, while the out-modes $u_{k'}^{\rm out}$ are chosen to be positive-frequency near $\Sigma_{\rm out}$. These choices specify the in- and out-vacua, and the two bases are related by the Bogoliubov transformation
$u_{k'}^{\rm out}
=
\sum_k
\left(
\alpha_{k'k}u_k^{\rm in}
+
\beta_{k'k}u_k^{{\rm in}*}
\right)$,
where $\alpha$ and $\beta$ are the Bogoliubov matrices. 
Throughout, we work in a consistently regularized theory, in which $\alpha$ is assumed to be invertible, all traces and determinants appearing below are well defined in the usual matrix sense, and the interchange of integrations and summations is justified. The regulator is removed only after all manipulations have been completed within the regularized theory.

The mode normalization is fixed by the Wronskian
\begin{align}
W_\Sigma[f,g]
=
\int_\Sigma \dd\Sigma_\mu
\left(
f\nabla^\mu g-g\nabla^\mu f
\right),
\end{align}
with $\dd\Sigma_{\mu}=\dd\Sigma\, n_{\mu}$, where $\dd \Sigma$ is the surface element on the Cauchy surface $\Sigma$, and $n^{\mu}$ is the future-directed unit normal to $\Sigma$. We use the convention
$W_\Sigma[u_k^{\rm in},u_j^{{\rm in}*}]
=
i\delta_{kj},
W_\Sigma[u_{k'}^{\rm out},u_{j'}^{{\rm out}*}]
=
i\delta_{k'j'}$,
with the Wronskians between two positive-frequency modes or two negative-frequency modes vanishing. Since the mode functions solve the homogeneous equation, these Wronskians are independent of the choices of the Cauchy surface.
The in-out Feynman Green's function can be expressed in the in-out bases as
$G_{\rm F}(x,x')
=\theta(x,x')
\sum_{k',k}
(\alpha^{-1})_{kk'}
u_{k'}^{\rm out}(x)u_k^{{\rm in}*}(x')
+ \theta(x',x)
\sum_{k',k}
(\alpha^{-1})_{kk'}
u_{k'}^{\rm out}(x') u_k^{{\rm in}*}(x).$
Here, the step function $\theta(x,x')$ indicates that $x$ lies in the future of $x'$.  With the symmetric prescription $\theta(x,x)=1/2$, the two branches contribute equally in the coincident limit, giving
\begin{align}
G_{\rm F}(x,x)
=
\sum_{k',k}
(\alpha^{-1})_{kk'}u_{k'}^{\rm out}(x)
u_k^{{\rm in}*}(x). 
\label{eq:GFcoincident}
\end{align}
A detailed derivation of the in-out mode representation of $G_{\rm F}$ is provided in the  Supplemental Material Sec.~II A \cite{supplemental}.

The bulk trace is therefore controlled by the integral of the form
$\int_\Omega \dd^{d+1}x\sqrt{|g|}\,u_{k'}^{\rm out}u_k^{{\rm in}*}$.
Since the modes satisfy the homogeneous equation, the Green–Lagrange identity converts this product into the divergence of a Wronskian current
$\nabla_\mu
\left(
u_{k'}^{\rm out}\nabla^\mu\partial_{m^2} u_k^{{\rm in}*}
-
\partial_{m^2}u_k^{{\rm in}*}\nabla^\mu u_{k'}^{\rm out}
\right)
=
u_{k'}^{\rm out}u_k^{{\rm in}*}$ \cite{baer2008waveequationslorentzianmanifolds}. 
Applying Gauss' theorem over $\Omega$, and assuming that the contribution from any lateral boundary vanishes, the integral becomes,
\begin{align}
\int_\Omega \dd^{d+1}x\sqrt{|g|}\,
u_{k'}^{\rm out}u_k^{{\rm in}*}
=&
W_{\Sigma_{\rm in}}
\left[
u_{k'}^{\rm out},\partial_{m^2} u_k^{{\rm in}*}
\right] \notag\\
&-
W_{\Sigma_{\rm out}}
\left[
u_{k'}^{\rm out},\partial_{m^2} u_k^{{\rm in}*}
\right].
\end{align}
in which only contributions from the initial and final Cauchy surfaces remain. The endpoint Wronskians can be reorganized using the overlap relation
$W_\Sigma[u_{k'}^{\rm out},u_k^{{\rm in}*}]=i\alpha_{k'k}$.
This allows the variation of the Bogoliubov coefficient $\alpha_{k'k}$ to be separated from the remaining endpoint contributions. The detailed reduction of the Green's function spacetime integral to a boundary Wronskian form is given in the Supplemental Material Sec.~II A \cite{supplemental}. Finally, the mass variation reconstructed from the Green's function prescription becomes
\begin{align}
\partial_{m^2} W_{\rm G} =&\frac{i}{2}\sum_{k,k'}(\alpha^{-1})_{kk'}\partial_{m^2}\alpha_{k'k} \notag\\
&-\frac{1}{2}\sum_{k,k'}(\alpha^{-1})_{kk'}W_{\Sigma_{\rm in}}\left[u_{k'}^{\rm out},\partial_{m^2}u_k^{{\rm in}*}\right]  \notag\\
&-\frac{1}{2}\sum_{k,k'}(\alpha^{-1})_{kk'}W_{\Sigma_{\rm out}}\left[\partial_{m^2}u_{k'}^{\rm out},u_k^{{\rm in}*}\right]
\end{align}
The first term on the right is exactly the mass variation of the Bogoliubov expression \eqref{eq:BogoliubovFormula}, while the remaining two terms involve the endpoint Wronskians.  Denoting them by $\mathcal B_{\rm in}$ and $\mathcal B_{\rm out}$, we obtain
\begin{align}
\partial_{m^2} W_G =\partial_{m^2}W_B-\mathcal{B}_{\rm in}-\mathcal{B}_{\rm out}.
\label{eq:dWG-obstruction}
\end{align}
The Green's function prescription therefore does not by itself reproduce the Bogoliubov variation: the remaining endpoint Wronskian terms constitute a boundary obstruction.

\paragraph{Boundary completion by vacuum wavefunctionals.}
We now evaluate the endpoint contribution from the vacuum wavefunctionals in the complete variation formula \eqref{eq:complete_variation}. This contribution is absent from the Green's function prescription and will be shown to cancel the boundary obstruction in \eqref{eq:dWG-obstruction}.

To evaluate the endpoint wavefunctional contributions, we introduce the boundary data of the mode functions. For either the in- or out-basis, denoted collectively by $\sigma=\{{\rm in},{\rm out}\}$, define on a Cauchy surface $\Sigma_{\sigma}$,
$q_k^\sigma(\bm x)
\equiv
u_k^\sigma\big|_{\Sigma_\sigma}(\bm x),
p_k^\sigma(\bm x)
\equiv
n^\mu\nabla_\mu u_k^\sigma\big|_{\Sigma_\sigma}(\bm x)$, where $\bm{x}\in \Sigma_{\sigma}$. 
The annihilation conditions defining the in- and out-vacua fix the endpoint wavefunctionals to be Gaussian \cite{Weinberg:1995mt,Long_1998,Long_1998II}. We write them in the compact form as
\begin{equation}
\begin{aligned}
\Psi_\sigma[\varphi^\sigma]
&=
\mathcal{N}_\sigma
\exp\left[
\frac{i}{2}
\varphi^\sigma\cdot K^\sigma\cdot\varphi^\sigma
\right],  \\
K^\sigma(\bm x,\bm y)
=&
\sum_k
(q^{-1})_k^{\sigma *}(\bm x)
p_k^{\sigma *}(\bm y),
\quad {\bm x},{\bm y}\in \Sigma_{\sigma} .
\end{aligned}
\label{eq:VacuumWaveFunctional}
\end{equation}
Here $\varphi^\sigma$ denotes a boundary field configuration on $\Sigma_\sigma$, i.e. the restriction of the integration field $\phi$ to that endpoint.  The dot denotes integration over the endpoint surface, for instance, $X^\sigma\cdot Y^\sigma=\int_{\Sigma_\sigma}\dd\Sigma_{\bm x}\,X^\sigma(\bm x)Y^\sigma(\bm x)$.  The quantity $q^{-1}$ denotes the dual basis to the boundary mode functions, satisfying $(q^{-1})^\sigma_k\cdot q_j^\sigma=\delta_{kj}$ and $ \sum_k  q_k^{\sigma}(\bm x)  (q^{-1})_k^{\sigma}(\bm y) =\delta_{\Sigma_{\sigma}}(\bm x,\bm y)$.  With this notation the kernel satisfies $K^\sigma\cdot q_j^{\sigma *}=p_j^{\sigma *}$, showing that the endpoint Gaussian is fixed by the choice of positive-frequency boundary data.

The normalization factor $\mathcal{N}_\sigma$ is part of the endpoint wavefunctional and therefore must be varied together with the Gaussian kernel. Its form follows from the normalization of the vacuum state. To express this factor in terms of a well-defined determinant, we work temporarily with a finite $N$-site regularization of the boundary field space. The normalization factor then takes the form
\begin{equation}
\begin{aligned}
\mathcal{N}_{\sigma}&=(2\pi)^{-\frac{N}{4}}\left[\det_{\Sigma_{\sigma}}{Q^{\sigma}}\right]^{-\frac{1}{4}}e^{i\theta^{\sigma}},  \\
Q^\sigma(\bm x,\bm y)
&=
\sum_k
q_k^\sigma(\bm x)q_k^{\sigma *}(\bm y),\quad {\bm x},{\bm y}\in \Sigma_{\sigma} .
\end{aligned}
\label{eq:VacuumNormalization}
\end{equation}
The determinant is taken over the regulated boundary field space. The normalization condition fixes only $|\mathcal N_\sigma|$, leaving an arbitrary real phase $\theta^\sigma$ associated with the endpoint-state convention. Accordingly,
$\left \langle\partial_{m^2}\log\Psi_\sigma \right \rangle
=
\partial_{m^2}\log \mathcal{N}_\sigma
+
\frac{i}{2}
\left \langle \varphi^\sigma\cdot
\partial_{m^2}K^\sigma
\cdot\varphi^\sigma \right \rangle.$ 
The first term on the right is independent of the boundary field configuration, whereas the second term probes the boundary two-point function of the integration field.

The variation of the normalization prefactor is immediate:
\begin{eqnarray}
\partial_{m^2}\log \mathcal{N}_{\sigma}=-\frac{1}{2}
\sum_k\re\left[\left(q^{-1}\right)_k^{\sigma}\cdot\partial_{m^2}q_k^\sigma
\right] +i\partial_{m^2}\theta^\sigma.~~ 
\end{eqnarray}
The explicit $(2\pi)^{-N/4}$ factor is independent of $m^2$ and drops out. For the complex-conjugate normalization factor at the out-endpoint, the modulus gives the same contribution, while the phase term changes sign.

It remains to compute the Gaussian-kernel contribution.  This requires the boundary two-point function of the path-integral field.  This is obtained by restricting the in-out Feynman Green's function to the endpoint surface. Using  $\theta(x,x)=1/2$, one finds
\begin{align}
\left \langle\varphi^\sigma(\bm x)\varphi^\sigma(\bm y) \right \rangle
=&\frac{1}{2}\sum_{k,k'}\left( \alpha^{-1} \right)_{kk'}\left[ u_{k'}^{\rm out}(\bm x)u_{k}^{{\rm in}* }(\bm y)\right. \notag \\
&\left.
+u_{k'}^{\rm out}(\bm y)u_{k}^{{\rm in}* }(\bm x)\right], \quad {\bm x},{\bm y}\in \Sigma_{\sigma} .
\end{align}

We can now evaluate the endpoint wavefunctional contributions explicitly. Here we only sketch the calculation, with the details given in the Supplemental Material Sec.~II B \cite{supplemental}. On the Cauchy surface $\Sigma_{\rm in}$, $u_k^{{\rm in}*}|_{\Sigma_{\rm in}}=q_k^{{\rm in}*}$. The contraction of $\partial_{m^2}K^{\rm in}$ with the boundary two-point function is therefore determined by the in-mode Cauchy data, $q_k^{{\rm in}*}$ and $p_k^{{\rm in}*}$, together with their mass variations and the values of the out-modes on $\Sigma_{\rm in}$. The in-out Wronskian relation $W_{\Sigma_{\rm in}}[u_{k'}^{\rm out},u_j^{{\rm in}*}]=i\alpha_{k'j}$ further allows the out-mode boundary values to be related to their normal derivatives. With these relations, the Gaussian-kernel contribution can be organized into the Wronskian $W_{\Sigma_{\rm in}}[u_{k'}^{\rm out},\partial_{m^2}u_k^{{\rm in}*}]$, together with a remaining boundary trace:
\begin{align}
\left\langle\varphi^{\rm in}\cdot\partial_{m^2}K^{\rm in}\cdot\varphi^{\rm in} \right \rangle
&=\sum_{kk'}(\alpha^{-1})_{kk'}W_{\Sigma_{\rm in}}\left[u_{k'}^{\rm out},\partial_{m^2}u_{k}^{{\rm in} *} \right]\notag\\
&-i\sum_k\left[ \left(q^{-1}\right)_k^{{\rm in}*}\cdot\partial_{m^2}q_k^{{\rm in}*} \right].
\end{align}
Adding the prefactor variation to this kernel contribution gives the full initial wavefunctional variation:
\begin{eqnarray}
&&\left\langle
\partial_{m^2}\log\Psi_{\rm in}\right\rangle
=
\frac{i}{2}\sum_{kk'}(\alpha^{-1})_{kk'}W_{\Sigma_{\rm in}}\left[u_{k'}^{\rm out},\partial_{m^2}u_{k}^{{\rm in} *} \right]  \notag\\
&&~~~~~~+\frac{i}{2}
\sum_k\im\left[ \left(q^{-1}\right)_k^{{\rm in}*}\cdot\partial_{m^2}q_k^{{\rm in}*} \right] 
+i\partial_{m^2}\theta^{\rm in}.
\end{eqnarray}
The final endpoint has similar structure, with the complex conjugation $\Psi_{\rm out}^*$:
\begin{eqnarray}
&&\left\langle
\partial_{m^2}\log\Psi^*_{\rm out}\right\rangle
=
\frac{i}{2}\sum_{kk'}(\alpha^{-1})_{kk'}W_{\Sigma_{\rm out}}\left[\partial_{m^2}u_{k'}^{\rm out},u_{k}^{{\rm in} *} \right]\notag\\
&&~~~~~+\frac{i}{2}
\sum_{k'} \im\left[ \left(q^{-1}\right)_{k'}^{{\rm out} }\cdot\partial_{m^2}q_{k'}^{{\rm out} }
\right]
-i\partial_{m^2}\theta^{\rm out}. 
\end{eqnarray}
In both endpoint variations, the right side in the first line is a Wronskian contribution, while the remaining terms  in the second line are purely imaginary. After multiplication by $-i$ in $W_\Psi$, the latter contribute only to the real part of the effective action. The two Wronskian terms are exactly the endpoint structures denoted by $\mathcal B_{\rm in}$ and $\mathcal B_{\rm out}$ in \eqref{eq:dWG-obstruction}. Therefore, the endpoint wavefunctionals contribute
\begin{align}
\partial_{m^2}W_{\Psi}=+\mathcal{B}_{\rm in}+\mathcal{B}_{\rm out} +\partial_{m^2}\Theta.
\end{align}
This is the boundary completion: the vacuum wavefunctionals reproduce the endpoint Wronskian terms arising from the Green's function prescription \eqref{eq:dWG-obstruction}, but with the opposite signs, thereby exactly cancel the boundary obstruction. 
The remaining term $\Theta$ is real and collects the non-Wronskian endpoint contributions. Inserting this endpoint contribution into the complete variation formula \eqref{eq:complete_variation}, together with the bulk relation \eqref{eq:dWG-obstruction}, it gives
\begin{align}
\partial_{m^2}W
=
\partial_{m^2}W_{\rm B}
+
\partial_{m^2}\Theta .
\label{eq:WPsi-boundary-completion}
\end{align}
Thus, the complete in-out variation equals the Bogoliubov variation up to a real phase convention. And, the Green's function prescription becomes equivalent to the Bogoliubov prescription only after the endpoint wavefunctionals are included.

Integrating over $m^2$, the real term $\Theta$ can affect only the real part of $W$, while an $m^2$-independent constant $W_0$ may remain. We fix $W_0$ by taking the large-mass limit as the reference condition. In this limit particle production is suppressed, $\alpha\alpha^\dagger\to I$, and hence $\im W_0=0$. The imaginary part of the complete effective action is therefore
\begin{align}
\im W= \frac{1}{4}\tr \log \left[\alpha\alpha^{\dagger}\right].
\label{eq:boundary-completed-ImW}
\end{align}
This is precisely the Bogoliubov result: once the endpoint wavefunctionals are included consistently, the imaginary part of the complete in-out effective action is given by the Bogoliubov expression. Therefore, the result \eqref{eq:boundary-completed-ImW} obtained from the boundary completion of the Green's function prescription resolves the previous discrepancy without relying on a particular spacetime background.

\paragraph{Discussion.}
The immediate consequence of \eqref{eq:boundary-completed-ImW} is that the vacuum persistence probability is not prescription dependent:
$P_{\rm vac}=e^{-2\im W}=e^{-2\im W_{\rm B}}$.
The residual endpoint phase affects only the real part of $W$, leaving the vacuum persistence probability unchanged. Without an additional phase convention or renormalization condition, however, $\re W$ is not uniquely defined because it can be shifted by the normalization phase of the endpoint wavefunctionals.

Our analysis also clarifies the prescription dependence found in de Sitter particle production \cite{Akhmedov:2024qvi,Zhou:2025jwm}.  In that de Sitter calculation \cite{Zhou:2025jwm}, the effective action is reconstructed solely from the mode-function representation of the coincident Green's function and is sensitive to how the late-time and momentum cutoffs are correlated \cite{Zhou:2025jwm}. From the present viewpoint, this sensitivity reflects how the endpoint Wronskian term in \eqref{eq:dWG-obstruction} is treated. The vacuum wavefunctional contribution carries the corresponding prescription dependence with the opposite sign. When the two contributions are combined within the same regularization scheme, the prescription dependence cancels as demonstrated explicitly in the Supplemental Material Sec.~III B \cite{supplemental}. The cutoff analysis therefore diagnoses a genuine boundary sensitivity of the bulk Green's function prescription, rather than an intrinsic ambiguity of the complete vacuum persistence probability.

This framework may also be viewed as a generalization of the cosmological analysis presented in Ref.~\cite{Akhmedov:2024qvi}. We extend the analysis to general globally hyperbolic spacetimes and identify the boundary-completion mechanism that reconciles the Green's function and Bogoliubov approaches. Since the completion holds at the regulated level, the case of a background field acting over an infinite time interval can be treated as limits of regularized problems while preserving an unambiguous vacuum persistence probability. More generally, boundary completion does not require asymptotic regions: the in- and out-vacua may instead be specified on arbitrary Cauchy surfaces through their wavefunctionals. Thus, the lack of asymptotic regions does not preclude the application of the in-out formalism; what is required is a complete and consistent specification of the boundary states.

The boundary-completion framework admits an extension to more general quadratic theories. The key ingredients underlying our derivation—the linear mode equations and the Gaussian structure of the vacuum wavefunctionals—remain available in such theories. For example, in the presence of a background electromagnetic field, this involves replacing the relevant derivative operators with their gauge-covariant counterparts. By contrast, for theories with nonlinear interactions, the equations of motion are no longer linear and the vacuum wavefunctionals are generally non-Gaussian. Our present results therefore do not apply directly to such theories, and their extension requires further investigation. A potentially viable approach would be to treat the interactions perturbatively. Nevertheless, the underlying structural point remains: the vacuum wavefunctionals are part of the in-out path integral and should be included consistently.

It is also instructive to contrast the present discussion with the Schwinger–Keldysh formalism. The in-out formalism describes transition amplitudes between specified states, whereas the Schwinger–Keldysh formalism computes expectation values in a given initial state \cite{Hu:2020luk}. In the latter, unitarity implies $Z_{\rm SK}[J,J]=1$ and hence $W_{\rm SK}[J,J]=0$. The initial state nevertheless remains explicit in the closed-time-path integral through the initial density matrix. If the initial state depends on the parameter being varied, its variation must therefore be included together with that of the bulk action; otherwise, the resulting variation need not preserve the unitarity relation $Z_{\rm SK}[J,J]=1$. In this sense, the need for a consistent treatment of boundary-state data persists also in the Schwinger–Keldysh formalism, although there is no independent final-state contribution.

It is worth noting that a possible relation between the real and imaginary parts of the effective action has been suggested within the in-out formalism. In particular, studies based on Bogoliubov coefficients have conjectured a one-to-one correspondence between vacuum polarization and vacuum persistence, supported by a number of exactly solvable backgrounds \cite{Kim_2017,Kim_2008,Kim:2016xvg}. Our present analysis makes explicit the structure of the full in-out effective action: in addition to the Bogoliubov-dependent part, its real part contains contributions from the endpoint wavefunctionals, while its imaginary part remains fixed by the Bogoliubov coefficients. This structure provides a setting in which the proposed correspondence can be further tested and its possible extension to the full boundary-completed effective action investigated. However, such a test necessarily requires additional input, since the phase convention for the endpoint states and the renormalization prescription must be specified to uniquely determine the real part of the effective action.

A further extension concerns systems with nontrivial spatial boundary conditions \cite{Karabali_2013,Govindarajan_2014}, such as fields in AdS spacetime \cite{Ishibashi_2004,Kakkar_2022,Bishwakarma:2026xhq} or background electric fields described in a static gauge \cite{Nikishov_2004,Latifi:2025fya}. In such cases, Eq.~\eqref{eq:dWG-obstruction} generally acquires additional Green–Lagrange terms from spatial boundaries. Moreover, if the spatial boundary conditions depend on the parameter being varied, their variation may provide additional contributions to the effective action. This suggests a possible spatial analogue of the boundary-completion mechanism identified here, whose precise formulation is left for future work.

\paragraph{Conclusion.}
We have established a boundary-completed reconstruction of the in-out effective action by consistently incorporating the contributions from bulk Green's function and the endpoint vacuum states. The resulting vacuum persistence probability agrees with the Bogoliubov result. The construction is independent of the particular spacetime background and does not require asymptotic in- and out-regions, with the vacuum states specified instead on general Cauchy surfaces. Our result therefore provides a framework for reconstructing the complete in-out effective action.

\begin{acknowledgments}
This work was partially supported by the National Natural Science Foundation of China (Grants No.12175008 and No.12675057).
\end{acknowledgments}

\bibliography{Ref}
 \bibliographystyle{elsarticle-num}


\clearpage

\onecolumngrid

\renewcommand{\theequation}{S\arabic{equation}}
\setcounter{equation}{0}

\begin{center}
{\large\bf Supplemental Material for\\
``Boundary Completion of Vacuum Persistence Probability''}

\end{center}

We first point out the conventions and setups in the main text and in this Supplemental Material. Later, we list the detailed derivations related to the evaluations in the main text. Finally, we present two illustrations of the boundary-completion mechanism.

\section{I. Conventions and Setup}
Throughout the Letter and this Supplemental Material, we use the natural units, i.e., \(\hbar=c=1\), and the mostly-plus metric signature. With this convention, the action for a free real scalar field can be written as
\begin{align}
S_m[\phi]=-\frac{1}{2}
\int_{\Omega}\dd^{d+1}x\sqrt{|g|}
\left[g^{\mu\nu}\nabla_\mu\phi\nabla_\nu\phi+m^2\phi^2\right],
\end{align}
where \(m\) is the mass of the scalar field.

We consider a general \((d+1)\)-dimensional globally hyperbolic spacetime region \(\Omega\). We assume that \(\Omega\) is bounded in the time direction by two Cauchy surfaces, denoted by \(\Sigma_{\rm in}\) and \(\Sigma_{\rm out}\). These two hypersurfaces serve as the initial and final boundaries of the in-out problem. Possible lateral boundary contributions will be assumed to vanish.

The classical equation of motion following from the action is
\begin{align}
{\rm P_m}\phi\equiv\left(-\nabla_{\mu}\nabla^{\mu}+m^2 \right)\phi=0.
\end{align}
The field operator and the mode functions below obey this equation.

To quantize the scalar field, we introduce two complete sets of solutions $\{u_k^{\rm in},u_k^{{\rm in}*}\}$ and $\{u_{k'}^{\rm out},u_{k'}^{{\rm out}*}\}$ of the field equation. These are the in- and out-mode bases. The in-modes $u_k^{\rm in}$ are chosen to be positive-frequency near $\Sigma_{\rm in}$, whereas the out-modes $u_{k'}^{\rm out}$ are chosen to be positive-frequency near $\Sigma_{\rm out}$. The field operator can be expanded in either basis as
\begin{align}
\phi= \sum_{k}\left[u^{\rm in}_k a^{\rm in}_k +u^{\rm in *}_k a_k^{\rm in \dagger}\right] = \sum_{k'}\left[ u^{\rm out}_{k'} a^{\rm out}_{k'} +u^{\rm out *}_{k'} a_{k'}^{\rm out \dagger} \right],
\end{align}
where the corresponding creation and annihilation operators obey the standard canonical commutation relations,
\begin{align}
&\left[a^{\rm in}_{j},a_{k}^{\rm in \dagger} \right]=\delta_{jk}, \quad \left[a^{\rm in}_{j},a_{k}^{\rm in } \right]=\left[a^{\rm in \dagger}_{j},a_{k}^{\rm in \dagger} \right]=0, \\
&\left[a^{\rm out}_{j'},a_{k'}^{\rm out \dagger} \right]=\delta_{j'k'}, \quad \left[a^{\rm out}_{j'},a_{k'}^{\rm out } \right]=\left[a^{\rm out \dagger}_{j'},a_{k'}^{\rm out \dagger} \right]=0.
\end{align}
The in- and out-vacua are defined by the annihilation operators associated with the corresponding mode bases,
\begin{align}
&a_k^{\rm in}\left | 0,\rm in  \right \rangle= 0, \quad \forall k \\ &a_{k'}^{\rm out}\left | 0,\rm out  \right \rangle  = 0 , \quad \forall  k'
\end{align}
Thus the two choices of positive-frequency modes specify, respectively, the initial and final vacuum states entering the in-out amplitude.

The Wronskian normalization of the mode functions follows from the canonical quantization condition. We define the Wronskian by
\begin{align}
W_{\Sigma}[f,g]=\int _{\Sigma}\dd \Sigma_{\mu } \left(f\nabla^{\mu}g-g\nabla^{\mu}f \right),
\end{align}
with $\dd\Sigma_{\mu}=\dd\Sigma\, n_{\mu}$, where $\dd \Sigma$ is the surface element on the Cauchy surface $\Sigma$, and $n^{\mu}$ is the future-directed unit normal to $\Sigma$. For both the in- and out-bases, the canonical quantization condition implies
\begin{align}
&W_\Sigma[u_k^{\rm in},u_j^{{\rm in }*}] =i\delta_{kj},\quad W_\Sigma[u_k^{\rm in},u_j^{{\rm in }}]=W_\Sigma[u_k^{{\rm in}*},u_j^{{\rm in }*}]=0.\\
&W_\Sigma[u_{k'}^{\rm out},u_{j'}^{{\rm out}*}] =i\delta_{k'j'}, \quad W_\Sigma[u_{k'}^{\rm out},u_{j'}^{{\rm out }}]=W_\Sigma[u_{k'}^{{\rm out}*},u_{j'}^{{\rm out }*}]=0.
\end{align}
Since the mode functions obey the equations of motion, the above Wronskians are conserved and hence independent of the choice of Cauchy surface.

The in- and out-modes are related by a Bogoliubov transformation. We use the convention
\begin{align}
u_{k'}^{\rm out} = \sum_{k} \left( \alpha_{k'k}u_k^{\rm in}+\beta_{k'k}u_k^{{\rm in}*} \right),
\end{align}
where $\alpha_{k'k}$ is the mode-preserving Bogoliubov coefficient, while $\beta_{k'k}$ is the mode-mixing coefficient between positive- and negative-frequency sectors.
With this convention, the Bogoliubov coefficients are equivalently characterized by the Wronskians
\begin{align}
&W_\Sigma[u_{k'}^{\rm out},u_k^{{\rm in}*}]=i\alpha_{k'k}, \\
&W_\Sigma[u_{k'}^{\rm out},u_k^{\rm in}]=-i\beta_{k'k}.
\end{align}
The Wronskian normalization of the in- and out-bases implies the Bogoliubov identities
\begin{align}
\sum_k\left(\alpha_{k'k}\alpha_{j'k}^*-\beta_{k'k}\beta_{j'k}^*
\right)=\delta_{k'j'},\quad
\sum_k\left(\alpha_{k'k}\beta_{j'k}-\beta_{k'k}\alpha_{j'k}\right)=0.\\
\sum_{k'}\left(\alpha_{k'k}^*\alpha_{k'j}-\beta_{k'k}\beta_{k'j}^*
\right)=\delta_{kj}, \quad \sum_{k'}\left(\alpha_{k'k}^*\beta_{k'j}-\beta_{k'k}\alpha_{k'j}^*
\right)=0.
\end{align}
The inverse Bogoliubov transformation relation is
\begin{align}
u_k^{\rm in}=\sum_{j'}
\left(\alpha_{j'k}^{*}u_{j'}^{\rm out}-\beta_{j'k}u_{j'}^{{\rm out}*}
\right).
\end{align}
The corresponding transformations of the creation and annihilation operators are
\begin{align}
&a_{k'}^{\rm out}=\sum_k\left(\alpha_{k'k}^*a_k^{\rm in}-\beta_{k'k}^*a_k^{{\rm in}\dagger} 
\right), \\
&a_k^{\rm in}=
\sum_{j'}\left(\alpha_{j'k}a_{j'}^{\rm out}+\beta_{j'k}^*a_{j'}^{{\rm out}\dagger}\right).
\label{BogoliubovRelation-a}
\end{align}

Throughout the following derivation, all mode sums, traces, determinants, and Bogoliubov matrices are understood in a common regulator, so that they may be treated as finite-dimensional matrices until the regulator is removed.

\section{II. Detailed Derivation}
\subsection{A. Green's Function Contribution}
The in-out Feynman Green’s function is
\begin{align}
G_{\rm F}(x,x')=\frac{\left \langle0,{\rm out} |T\phi(x)\phi(x')| 0,{\rm in} \right \rangle }{\left \langle0,{\rm out} | 0,{\rm in} \right \rangle}.
\end{align}
Equivalently, separating the two time orderings, we have
\begin{align}
G_{\rm F}(x,x')=\theta(x,x')\frac{\left \langle0,{\rm out} |\phi(x)\phi(x')| 0,{\rm in} \right \rangle }{\left \langle0,{\rm out} | 0,{\rm in} \right \rangle}+\theta(x',x)\frac{\left \langle0,{\rm out} |\phi(x')\phi(x)| 0,{\rm in} \right \rangle }{\left \langle0,{\rm out} | 0,{\rm in} \right \rangle}.
\end{align}
Here $\theta(x,x')$ equals one when $x$ lies in the future of $x'$, and zero otherwise. We take $\theta(x,x)=1/2$ in the coincident limit.

For the first time ordering, the out-basis is used for $\phi(x)$ and the in-basis for $\phi(x')$. Since $a_{k'}^{{\rm out}\dagger} $ annihilates the out-bra and $a_k^{\rm in}$ annihilates the in-ket, the only nonvanishing contraction is
\begin{align}
\frac{\left \langle0,{\rm out} |\phi(x)\phi(x')| 0,{\rm in} \right \rangle }{\left \langle 0,{\rm out} | 0,{\rm in} \right \rangle}
=\sum_{k,k'}u_{k'}^{\rm out}(x)u_{k}^{{\rm in}* }(x')\frac{\left \langle0,{\rm out} \left|a^{{\rm out} }_{k'}a_k^{{\rm in}\dagger}\right| 0,{\rm in} \right \rangle }{\left \langle0,{\rm out} | 0,{\rm in} \right \rangle}.
\end{align}
The mixed in-out contraction is
\begin{align}
\frac{\left \langle0,{\rm out} \left|a^{{\rm out} }_{k'}a_k^{{\rm in}\dagger}\right| 0,{\rm in} \right \rangle }{\left \langle0,{\rm out} | 0,{\rm in} \right \rangle} = \left( \alpha^{-1} \right)_{kk'},
\end{align}
where $\alpha^{-1}$ denotes the inverse of $\alpha$, satisfying $ \sum_{k'}\left( \alpha^{-1} \right)_{jk'} \alpha_{k'k}=\delta_{jk}$ and $\sum_k\alpha_{j'k} \left( \alpha^{-1} \right)_{kk'} =\delta_{j'k'}$. 
This contraction can be checked directly from the Bogoliubov transformation \eqref{BogoliubovRelation-a}
\begin{align}
\sum_k\alpha_{j'k}\frac{\left \langle0,{\rm out} \left|a^{{\rm out} }_{k'}a_k^{{\rm in}\dagger}\right| 0,{\rm in} \right \rangle }{\left \langle0,{\rm out} | 0,{\rm in} \right \rangle} 
&=\frac{\left \langle0,{\rm out} \left|a^{{\rm out} }_{k'}\sum_k\left(\alpha_{j'k}a_k^{{\rm in}\dagger}-\beta_{j'k}a_k^{\rm in}\right)\right| 0,{\rm in} \right \rangle }{\left \langle0,{\rm out} | 0,{\rm in} \right \rangle} \notag\\
&=\frac{\left \langle0,{\rm out} \left|a^{{\rm out} }_{k'}a_{j'}^{{\rm out}\dagger}\right| 0,{\rm in} \right \rangle }{\left \langle0,{\rm out} | 0,{\rm in} \right \rangle}=\delta_{j'k'}.
\end{align}
This verifies the right-inverse relation. The left-inverse relation can be checked in the same way by expanding $a_j^{\rm in}$ in terms of out operators.

We then obtain
\begin{align}
\frac{\left \langle0,{\rm out} |\phi(x)\phi(x')| 0,{\rm in} \right \rangle }{\left \langle 0,{\rm out} | 0,{\rm in} \right \rangle}
=\sum_{k,k'}\left( \alpha^{-1} \right)_{kk'}u_{k'}^{\rm out}(x)u_{k}^{{\rm in}* }(x').
\end{align}
The second time ordering is obtained by interchanging $x$ and $x'$. Therefore the in-out Feynman Green's function takes the form
\begin{align}
G_{\rm F}(x,x')=\theta(x,x')\sum_{k,k'} \left( \alpha^{-1} \right)_{kk'}u_{k'}^{\rm out}(x)u_{k}^{{\rm in}* }(x')
+\theta(x',x)\sum_{k,k'}\left( \alpha^{-1} \right)_{kk'}u_{k'}^{\rm out}(x')u_{k}^{{\rm in}* }(x).
\end{align}
Taking the coincident limit and using the prescription $\theta(x,x)=1/2$, the two time-ordered branches give identical contributions. Therefore,
\begin{align}
G_{\rm F}(x,x)=
\sum_{k',k}\left(\alpha^{-1}\right)_{kk'}u_{k'}^{\rm out}(x)u_k^{{\rm in}*}(x).
\end{align}

Substituting this expression into the bulk mass variation reduces the Green's function contribution to spacetime integrals of the form
\begin{align}
\int_{\Omega}\dd^{d+1}x\sqrt{|g|}u_{k'}^{\rm out}(x)u_k^{{\rm in}*}(x).
\end{align}
The integrand can be written as a total divergence by means of the Green–Lagrange identity
\begin{align}
\nabla_\mu
\left( u_{k'}^{\rm out}\nabla^\mu\partial_{m^2} u_k^{{\rm in}*}
-\partial_{m^2}u_k^{{\rm in}*}\nabla^\mu u_{k'}^{\rm out}
\right)
=u_{k'}^{\rm out}u_k^{{\rm in}*}.
\end{align}
More generally, for any two scalar functions $f$ and $g$, the Green–Lagrange identity reads
\begin{align}
\nabla_\mu
\left(f\nabla^\mu g-g\nabla^\mu f
\right)= g{\rm P_m} f-f{\rm P_m}  g,
\qquad
\text{where}~{\rm P_m} =-\nabla_\mu\nabla^\mu+m^2 .
\end{align}
Taking $f=u_{k'}^{\rm out}$ and $g=\partial_{m^2}u_k^{{\rm in}*}$, and using
\begin{align}
{\rm P_m} u_{k'}^{\rm out}=0,
\qquad
{\rm P_m}\partial_{m^2}u_k^{{\rm in}*}=
-u_k^{{\rm in}*},
\end{align}
one obtains
\begin{align}
\nabla_\mu\left(u_{k'}^{\rm out}\nabla^\mu\partial_{m^2}u_k^{{\rm in}*}-
\partial_{m^2}u_k^{{\rm in}*}\nabla^\mu u_{k'}^{\rm out}\right)
=u_{k'}^{\rm out}u_k^{{\rm in}*}.
\end{align}
Integrating this identity over the spacetime region $\Omega$ gives
\begin{align}
\int_{\Omega}\dd^{d+1}x\sqrt{|g|}u_{k'}^{\rm out}(x)u_k^{{\rm in}*}(x)=\int_{\Omega}\dd^{d+1}x\sqrt{|g|}\nabla_\mu\left(u_{k'}^{\rm out}\nabla^\mu\partial_{m^2}u_k^{{\rm in}*}-
\partial_{m^2}u_k^{{\rm in}*}\nabla^\mu u_{k'}^{\rm out}\right).
\end{align}
We use Gauss' theorem in Lorentzian signature in the form
\begin{align}
\int_{\Omega}\dd^{d+1}x\sqrt{|g|} \nabla_\mu J^\mu=
\int_{\partial\Omega}\dd\Sigma\,\epsilon_{\bar n} \bar{n}_\mu J^\mu ,
\label{eq:gauss-lorentzian}
\end{align}
where $\bar{n}^\mu$ is the outward-pointing unit normal to $\partial\Omega$, and $\epsilon_{\bar{n}}=\bar{n}^\mu \bar{n}_\mu=\pm 1 $. We now apply \eqref{eq:gauss-lorentzian} to $J^{\mu}=u_{k'}^{\rm out}\nabla^\mu\partial_{m^2}u_k^{{\rm in}*}-\partial_{m^2}u_k^{{\rm in}*}\nabla^\mu u_{k'}^{\rm out}$. We assume that the lateral boundary contribution vanishes, so only the contributions from the initial and final Cauchy surfaces remain. Since these Cauchy surfaces are spacelike, their unit normal satisfies $\epsilon_{\bar{n}}=-1$. Hence
\begin{align}
\int_{\Omega}\dd^{d+1}x\sqrt{|g|}u_{k'}^{\rm out}(x)u_k^{{\rm in}*}(x)=-\int_{\Sigma_{\rm out}}\dd \Sigma \bar{n}_{\mu}J^{\mu}-\int_{\Sigma_{\rm in}}\dd \Sigma \bar{n}_{\mu}J^{\mu}.
\end{align}
On the final and initial Cauchy surfaces the outward-pointing normal is related to the future-directed normal by
\begin{align}
\bar n^\mu\big|_{\Sigma_{\rm out}} =n^\mu\big|_{\Sigma_{\rm out}},
\qquad
\bar n^\mu\big|_{\Sigma_{\rm in}} =- n^\mu\big|_{\Sigma_{\rm in}} .
\end{align}
Therefore,
\begin{align}
\int_{\Omega}\dd^{d+1}x\sqrt{|g|}\,
u_{k'}^{\rm out}(x)u_k^{{\rm in}*}(x)
&=
\int_{\Sigma_{\rm in}}\dd\Sigma\, n_\mu J^\mu
-
\int_{\Sigma_{\rm out}}\dd\Sigma\, n_\mu J^\mu
\nonumber\\
&=
W_{\Sigma_{\rm in}}
\left[
u_{k'}^{\rm out},
\partial_{m^2}u_k^{{\rm in}*}
\right]
-
W_{\Sigma_{\rm out}}
\left[
u_{k'}^{\rm out},
\partial_{m^2}u_k^{{\rm in}*}
\right].
\end{align}
To isolate the Bogoliubov contribution, we use the differentiated Wronskian relation on the final Cauchy surface. Since
$W_{\Sigma_{\rm out}}\left[u_{k'}^{\rm out},u_k^{{\rm in}*} \right]=i\alpha_{k'k}$, differentiation with respect to $m^2$ gives
\begin{align}
W_{\Sigma_{\rm out}}
\left[
u_{k'}^{\rm out},
\partial_{m^2}u_k^{{\rm in}*}
\right]=
i\partial_{m^2}\alpha_{k'k}
-
W_{\Sigma_{\rm out}}
\left[
\partial_{m^2}u_{k'}^{\rm out},
u_k^{{\rm in}*}
\right].
\end{align}
Substituting this into the boundary expression for the spacetime integral, we obtain
\begin{align}
\int_{\Omega}\dd^{d+1}x\sqrt{|g|}
u_{k'}^{\rm out}u_k^{{\rm in}*}
&=
-i\partial_{m^2}\alpha_{k'k}
+
W_{\Sigma_{\rm in}}
\left[
u_{k'}^{\rm out},
\partial_{m^2}u_k^{{\rm in}*}
\right]
+
W_{\Sigma_{\rm out}}
\left[
\partial_{m^2}u_{k'}^{\rm out},
u_k^{{\rm in}*}
\right].
\end{align}
Substitution into the bulk Green's function contribution $\partial_{m^2} W_G = -\frac{1}{2}\int_{\Omega}\dd ^{d+1}x\sqrt{|g|}G_F(x,x)$ gives
\begin{align}
\partial_{m^2} W_{\rm G} =&\frac{i}{2}\sum_{k,k'}(\alpha^{-1})_{kk'}\partial_{m^2}\alpha_{k'k} \notag\\
&-\frac{1}{2}\sum_{k,k'}(\alpha^{-1})_{kk'}W_{\Sigma_{\rm in}}\left[u_{k'}^{\rm out},\partial_{m^2}u_k^{{\rm in}*}\right] 
-\frac{1}{2}\sum_{k,k'}(\alpha^{-1})_{kk'}W_{\Sigma_{\rm out}}\left[\partial_{m^2}u_{k'}^{\rm out},u_k^{{\rm in}*}\right] .
\label{eq.WG}
\end{align}
The first term is the mass variation of the Bogoliubov expression. The remaining two terms are endpoint Wronskian contributions from the bulk Green's function variation. These terms will be combined with the endpoint wavefunctional contribution below.

\subsection{B. Vacuum Wavefunctional Contribution}
The endpoint wavefunctional contribution to the mass variation is
\begin{align}
\partial_{m^2}W_{\Psi}=-i
\left\langle
\partial_{m^2}\log\Psi_{\rm in}
\right\rangle
-i
\left\langle
\partial_{m^2}\log\Psi_{\rm out}^{*}
\right\rangle.
\label{eq.W-Psi}
\end{align}
The vacuum wavefunctionals are written in terms of the boundary data of the mode functions, which are defined as follows.

Fix an endpoint Cauchy surface $\Sigma_\sigma$, with $\sigma=\{{\rm in},{\rm out}\}$. Boundary coordinates on $\Sigma_\sigma$ are denoted by $\bm x,\bm y,\bm z$, while mode labels are denoted by $j,k,\ell$. The boundary data of the mode functions are defined by
\begin{align}
q_k^\sigma(\bm x)
\equiv
u_k^\sigma\big|_{\Sigma_\sigma}(\bm x),
\quad
p_k^\sigma(\bm x)
\equiv
n^\mu\nabla_\mu u_k^\sigma\big|_{\Sigma_\sigma}(\bm x), \quad \bm{x}\in \Sigma_{\sigma}.
\end{align}
Here $q_k^\sigma$ and $p_k^\sigma$ denote, respectively, the boundary value and the normal derivative of the mode function on $\Sigma_\sigma$. The corresponding boundary data of the complex-conjugate modes are denoted by $q_k^{\sigma *}(\bm x)$ and $p_k^{\sigma *}(\bm x)$. 

The Schrödinger vacuum wavefunctional associated with the $\sigma$-basis is fixed by the annihilation conditions
$a_k^\sigma\Psi_\sigma[\varphi^\sigma]=0$. Denote the boundary field and   momentum by $\varphi^\sigma(\bm x)=\phi|_{\Sigma_\sigma}(\bm x)$ and
$\pi^\sigma(\bm x)=n^\mu\nabla_\mu\phi|_{\Sigma_\sigma}(\bm x)$.
The annihilation operator associated with the $\sigma$-basis can be written as the Wronskian projection
\begin{align}
a_k^\sigma
=
iW_{\Sigma_\sigma}
\left[
u_k^{\sigma *},
\phi
\right]
=
i\int_{\Sigma_\sigma}\dd\Sigma_{\bm x}
\left[
q_k^{\sigma *}(\bm x)\pi^\sigma(\bm x)
-
p_k^{\sigma *}(\bm x)\varphi^\sigma(\bm x)
\right].
\end{align}
In the Schrödinger representation,
$\pi^\sigma(\bm x)=-i\delta/\delta\varphi^\sigma(\bm x)$. Therefore the vacuum condition becomes
\begin{align}
\int_{\Sigma_\sigma}\dd\Sigma_{\bm x}\,
q_k^{\sigma *}(\bm x)
\frac{\delta\Psi_\sigma}{\delta\varphi^\sigma(\bm x)}
=
i\int_{\Sigma_\sigma}\dd\Sigma_{\bm x}\,
p_k^{\sigma *}(\bm x)\varphi^\sigma(\bm x)
\Psi_\sigma .
\label{eq:annihilation-functional-equation}
\end{align}

The solution of \eqref{eq:annihilation-functional-equation} is taken in the
Gaussian form
\begin{align}
\Psi_\sigma[\varphi^\sigma]
=
\mathcal{N}_\sigma
\exp\left[
\frac{i}{2}
\int_{\Sigma_\sigma}\dd\Sigma_{\bm x}
\int_{\Sigma_\sigma}\dd\Sigma_{\bm y}\,
\varphi^\sigma(\bm x)
K^{\sigma}(\bm x,\bm y)
\varphi^\sigma(\bm y)
\right].
\label{eq:Psi-sigma-gaussian}
\end{align}
The kernel is chosen symmetric,
\begin{align}
K^{\sigma}(\bm x,\bm y)=K^{\sigma}(\bm y,\bm x),
\end{align}
so that
\begin{align}
\frac{\delta\Psi_\sigma}{\delta\varphi^\sigma(\bm x)}
=
i\int_{\Sigma_\sigma}\dd\Sigma_{\bm y}\,
K^{\sigma}(\bm x,\bm y)\varphi^\sigma(\bm y)\Psi_\sigma .
\end{align}
Substituting this into \eqref{eq:annihilation-functional-equation} gives
\begin{align}
\int_{\Sigma_\sigma}\dd\Sigma_{\bm x}\,
q_k^{\sigma *}(\bm x)
K^{\sigma}(\bm x,\bm y)
=
p_k^{\sigma *}(\bm y).
\label{eq:kernel-equation}
\end{align}
To invert this relation, we assume that the boundary functions $q_k^\sigma(\bm x)$ define a nondegenerate basis in the boundary field space. We then introduce the dual functions $(q^{-1})_k^\sigma(\bm x)$, defined by
\begin{align}
\int_{\Sigma_\sigma}\dd\Sigma_{\bm x}\,
(q^{-1})_j^\sigma(\bm x)q_k^\sigma(\bm x)
&=
\delta_{jk},
\label{eq:q-inverse-dual}
\\
\sum_k
q_k^\sigma(\bm x)(q^{-1})_k^\sigma(\bm y)
&=
\delta_{\Sigma_\sigma}(\bm x,\bm y).
\label{eq:q-inverse-complete}
\end{align}
Their complex conjugates satisfy
\begin{align}
\int_{\Sigma_\sigma}\dd\Sigma_{\bm x}\,
(q^{-1})_j^{\sigma *}(\bm x)q_k^{\sigma *}(\bm x)
&=
\delta_{jk},
\label{eq:qstar-inverse-dual}
\\
\sum_k
q_k^{\sigma *}(\bm x)(q^{-1})_k^{\sigma *}(\bm y)
&=
\delta_{\Sigma_\sigma}(\bm x,\bm y).
\label{eq:qstar-inverse-complete}
\end{align}
Using these relations, the solution of \eqref{eq:kernel-equation} is
\begin{align}
K^{\sigma}(\bm x,\bm y)
=
\sum_k
(q^{-1})_k^{\sigma *}(\bm x)
p_k^{\sigma *}(\bm y).
\label{eq:Ksigma-coordinate}
\end{align}
This can be verified by substituting \eqref{eq:Ksigma-coordinate} back into \eqref{eq:kernel-equation}. Using \eqref{eq:qstar-inverse-dual}, one finds
\begin{align}
\int_{\Sigma_\sigma}\dd\Sigma_{\bm x}\,
q_j^{\sigma *}(\bm x)
K^{\sigma}(\bm x,\bm y)
&=
\sum_k
\left[
\int_{\Sigma_\sigma}\dd\Sigma_{\bm x}\,
q_j^{\sigma *}(\bm x)
(q^{-1})_k^{\sigma *}(\bm x)
\right]
p_k^{\sigma *}(\bm y)
\nonumber\\
&=
p_j^{\sigma *}(\bm y).
\end{align}
The symmetry \(K^{\sigma}(\bm x,\bm y)=K^{\sigma}(\bm y,\bm x)\) follows from the vanishing Wronskians between two negative-frequency modes,
$W_{\Sigma_\sigma}\left[u_j^{\sigma *},u_k^{\sigma *}\right]=0$.

The normalization is fixed by the norm of the Schrödinger wavefunctional 
\begin{align}
\int_{\Sigma_{\sigma}}\mathcal{D}\varphi \Psi^*_\sigma[\varphi^\sigma]\Psi_\sigma[\varphi^\sigma] =1.
\end{align}
Since the boundary field configuration is real, the Gaussian wavefunctional gives
\begin{align}
|\mathcal{N}_{\sigma}|^2\int_{\Sigma_{\sigma}}\mathcal{D}\varphi \exp\left[
-\int_{\Sigma_\sigma}\dd\Sigma_{\bm x}
\int_{\Sigma_\sigma}\dd\Sigma_{\bm y}\,
\varphi^\sigma(\bm x)
\im K^{\sigma}(\bm x,\bm y)
\varphi^\sigma(\bm y)
\right]=1.
\end{align}
In a finite-$N$ lattice regulator on $\Sigma_\sigma$, the Gaussian integral gives
\begin{align}
|\mathcal{N}_{\sigma}|^2 = \pi^{-\frac{N}{2}}\left[\det_{\Sigma_{\sigma}}{\im K}\right]^{\frac{1}{2}}.
\label{eq:Nmod-ImK}
\end{align}
We further simplify $\im K^{\sigma}$ for later use. Using the Wronskian normalization $W_{\Sigma_{\sigma}}\left[u_j^{\sigma *},u_k^\sigma\right]=-i\delta_{jk}$, we multiply both sides by $(q^{-1})_j^{\sigma *}(\bm x)(q^{-1})_k^\sigma(\bm y)$ and sum over $j,k$. This gives
\begin{align}
K^{\sigma}(\bm x,\bm y)-K^{\sigma *}(\bm x,\bm y) =i\sum_k(q^{-1})_k^{\sigma *}(\bm x)(q^{-1})_k^\sigma(\bm y),
\label{eq:KminusKstar}
\end{align}
where the symmetry of $K^{\sigma}$ has been used. The sum on the right-hand side is the inverse of the boundary kernel
\begin{align}
Q^\sigma(\bm x,\bm y)
=
\sum_k
q_k^\sigma(\bm x)q_k^{\sigma *}(\bm y).
\label{eq:Qsigma}
\end{align}
Using the duality and completeness relations of $(q^{-1})_k^\sigma$,
one has
\begin{align}
\left(Q^{-1}\right)^{\sigma}(\bm x,\bm y)
=
\sum_k
(q^{-1})_k^{\sigma *}(\bm x)(q^{-1})_k^\sigma(\bm y).
\label{eq:Qinv}
\end{align}
To verify this, compute
\begin{align}
&
\int_{\Sigma_\sigma}\dd\Sigma_{\bm z}\,
Q_\sigma(\bm x,\bm z)
\sum_j
(q^{-1})_j^{\sigma *}(\bm z)(q^{-1})_j^\sigma(\bm y)
\nonumber\\
&\quad
=
\sum_{k,j}
q_k^\sigma(\bm x)(q^{-1})_j^\sigma(\bm y)
\int_{\Sigma_\sigma}\dd\Sigma_{\bm z}\,
q_k^{\sigma *}(\bm z)(q^{-1})_j^{\sigma *}(\bm z)
\nonumber\\
&\quad
=
\sum_k
q_k^\sigma(\bm x)(q^{-1})_k^\sigma(\bm y)
=
\delta_{\Sigma_\sigma}(\bm x,\bm y).
\end{align}
The right-inverse relation follows in the same way.

Therefore \eqref{eq:KminusKstar} becomes
\begin{align}
K^{\sigma}(\bm x,\bm y)-K^{\sigma *}(\bm x,\bm y)
=
i \left(Q^{-1}\right)^{\sigma}(\bm x,\bm y),
\end{align}
and hence
\begin{align}
\im K^{\sigma}(\bm x,\bm y)
=
\frac{1}{2}
\left(Q^{-1}\right)^{\sigma}(\bm x,\bm y).
\label{eq:ImK}
\end{align}
Using this relation, the normalization coefficient can be written entirely in
terms of the boundary kernel $Q_\sigma$. Substituting \eqref{eq:ImK} into \eqref{eq:Nmod-ImK}, one obtains
\begin{align}
|\mathcal{N}_{\sigma}|^2 = (2\pi)^{-\frac{N}{2}}\left[\det_{\Sigma_{\sigma}}{\left(Q^{-1}\right)^{\sigma}}\right]^{\frac{1}{2}}=(2\pi)^{-\frac{N}{2}}\left[\det_{\Sigma_{\sigma}}{Q^{\sigma}}\right]^{-\frac{1}{2}}.
\end{align}
In the second equality we used that, in the same finite-dimensional regulator, the inverse kernels satisfy $\det_{\Sigma_\sigma}Q_\sigma^{-1}= \left[\det_{\Sigma_\sigma}Q_\sigma\right]^{-1}$. The normalization condition fixes only the modulus of $\mathcal N_\sigma$. The normalization constant may still contain an arbitrary phase, which we write as
\begin{align}
\mathcal{N}_{\sigma}=(2\pi)^{-\frac{N}{4}}\left[\det_{\Sigma_{\sigma}}{Q^{\sigma}}\right]^{-\frac{1}{4}}e^{i\theta^{\sigma}}.
\end{align}
Here $\theta^{\sigma}$ denotes an arbitrary real phase that parametrizes the remaining phase convention.

The mass variation of the Gaussian wavefunctional separates into the variation of the normalization factor and the variation of the Gaussian kernel.
\begin{align}
\partial_{m^2}\log \Psi_{\sigma} =  \partial_{m^2}\log \mathcal{N}_{\sigma}
+\frac{i}{2}\int_{\Sigma_\sigma}\dd\Sigma_{\bm x}
\int_{\Sigma_\sigma}\dd\Sigma_{\bm y}\,
\varphi^\sigma(\bm x)
\partial_{m^2}K^{\sigma}(\bm x,\bm y)
\varphi^\sigma(\bm y),
\end{align}
\begin{align}
\partial_{m^2}\log \Psi^*_{\sigma} =  \partial_{m^2}\log \mathcal{N}^*_{\sigma}
-\frac{i}{2}\int_{\Sigma_\sigma}\dd\Sigma_{\bm x}
\int_{\Sigma_\sigma}\dd\Sigma_{\bm y}\,
\varphi^\sigma(\bm x)
\partial_{m^2}K^{\sigma*}(\bm x,\bm y)
\varphi^\sigma(\bm y).
\end{align}
We first evaluate the normalization contribution. Using the normalization constant obtained above, the $m^2$-independent factor drops out, while the arbitrary phase gives an additional phase-variation term.
\begin{align}
 \partial_{m^2}\log \mathcal{N}_{\sigma} &= \partial_{m^2}\log\left[\det_{\Sigma_{\sigma}}{Q^{\sigma}}\right]^{-\frac{1}{4}}+i\partial_{m^2}\theta^{\sigma} \\
 &=-\frac{1}{4}\partial_{m^2}\tr_{\Sigma_{\sigma}}\log Q^{\sigma}+i\partial_{m^2}\theta^{\sigma} \\
 &=-\frac{1}{4}\tr_{\Sigma_{\sigma}}\left[ Q^{-1}\partial_{m^2}Q \right]+i\partial_{m^2}\theta^{\sigma}.
\end{align}
The trace here is taken over the regulated boundary field space. In coordinate notation it is written as follows
\begin{align}
\tr_{\Sigma_{\sigma}}\left[ Q^{-1}\partial_{m^2}Q \right] &= \int_{\Sigma_{\sigma}}\dd\Sigma_{\bm x}\int_{\Sigma_{\sigma}}\dd\Sigma_{\bm z} Q^{-1}(\bm x,\bm z)\partial_{m^2}Q(\bm z,\bm x) \\
&=\sum_k\int_{\Sigma_\sigma}\dd \Sigma_{\bm x}\left[ \left(q^{-1}\right)^{\sigma}_{k}(\bm x)\partial_{m^2}q^{\sigma}_k(\bm x)
+\left(q^{-1}\right)^{\sigma*}_{k}(\bm x)\partial_{m^2}q^{\sigma*}_k(\bm x)\right].
\end{align}
Thus, the mass variation of the normalization factor takes the form
\begin{align}
\partial_{m^2}\log \mathcal{N}_{\sigma}=&-\frac{1}{4}\left[
\sum_k\int_{\Sigma_\sigma}\dd\Sigma_{\bm x} \left(q^{-1}\right)^{\sigma}_{k}(\bm x)\partial_{m^2}q_k^\sigma(\bm x)
+
\sum_k\int_{\Sigma_\sigma}\dd\Sigma_{\bm x}  \left(q^{-1}\right)^{\sigma *}_{k}(\bm x)\partial_{m^2}q_k^{\sigma *}(\bm x)
\right] \notag \\
&+i\partial_{m^2}\theta^{\sigma}.
\end{align}
For the complex-conjugate normalization factor, the modulus gives the same contribution, while the arbitrary phase changes sign.
\begin{align}
\partial_{m^2}\log \mathcal{N}^*_{\sigma}=&-\frac{1}{4}\left[
\sum_{k'}\int_{\Sigma_\sigma}\dd\Sigma_{\bm x} \left(q^{-1}\right)^{\sigma}_{k'}(\bm x)\partial_{m^2}q_{k'}^\sigma(\bm x)
+
\sum_{k'}\int_{\Sigma_\sigma}\dd\Sigma_{\bm x}  \left(q^{-1}\right)^{\sigma *}_{k'}(\bm x)\partial_{m^2}q_{k'}^{\sigma *}(\bm x)
\right] \notag \\
&-i\partial_{m^2}\theta^{\sigma}.
\end{align}
Since the normalization factors are independent of the boundary field configuration, their mass variations can be taken outside the path-integral expectation value. Therefore their contribution to $\langle \partial_{m^2}\log\Psi_\sigma\rangle$ is simply given by the expressions above. 

It remains to evaluate the contribution from the variation of the Gaussian kernel. This term contains the boundary field variables and
must be computed inside the expectation value. The following compact notation will be used for the quadratic boundary-field insertion.
\begin{align}
\frac{i}{2}\left \langle \varphi_{{\rm in}}\cdot \partial_{m^2}K^{\rm in}\cdot\varphi^{\rm in} \right \rangle
=&\left \langle \frac{i}{2}\int_{\Sigma_{\rm in}}\dd\Sigma_{\bm x}
\int_{\Sigma_{\rm in}}\dd\Sigma_{\bm y}\,
\varphi^{\rm in}(\bm x)
\partial_{m^2}K^{\rm in}(\bm x,\bm y)
\varphi^{\rm in}(\bm y) \right \rangle \notag\\
= &\frac{i}{2}\int_{\Sigma_{\rm in}}\dd\Sigma_{\bm x}
\int_{\Sigma_{\rm in}}\dd\Sigma_{\bm y}\,\partial_{m^2}K^{\rm in}(\bm x,\bm y)\left \langle\varphi^{\rm in}(\bm x)\varphi^{\rm in}(\bm y) \right \rangle,
\end{align}
\begin{align}
-\frac{i}{2}\left \langle \varphi_{{\rm out}}\cdot \partial_{m^2}K^{{\rm out}*}\cdot\varphi^{\rm out} \right \rangle
=&\left \langle -\frac{i}{2}\int_{\Sigma_{\rm out}}\dd\Sigma_{\bm x}
\int_{\Sigma_{\rm out}}\dd\Sigma_{\bm y}\,
\varphi^{\rm out}(\bm x)
\partial_{m^2}K^{{\rm out}*}(\bm x,\bm y)
\varphi^{\rm out}(\bm y) \right \rangle \notag\\
= &-\frac{i}{2}\int_{\Sigma_{\rm out}}\dd\Sigma_{\bm x}
\int_{\Sigma_{\rm out}}\dd\Sigma_{\bm y}\,\partial_{m^2}K^{{\rm out}*}(\bm x,\bm y)\left \langle\varphi^{\rm out}(\bm x)\varphi^{\rm out}(\bm y) \right \rangle.
\end{align}
The boundary two-point function is obtained by restricting the in-out Feynman Green's function to the corresponding endpoint Cauchy surface. With the symmetric coincident prescription, this gives
\begin{align}
\left \langle\varphi^\sigma(\bm x)\varphi^\sigma(\bm y) \right \rangle=\frac{1}{2}\sum_{k,k'}\left( \alpha^{-1} \right)_{kk'}\left[ u_{k'}^{\rm out}(\bm x)u_{k}^{{\rm in}* }(\bm y)
+u_{k'}^{\rm out}(\bm y)u_{k}^{{\rm in}* }(\bm x)\right],\quad {\bm x},{\bm y}\in\Sigma_{\sigma}.
\end{align}
The contribution from the in/out-vacuum wavefunctional can now be evaluated explicitly. Since the two calculations are completely parallel, we give the derivation for the in-vacuum wavefunctional and state the corresponding result for the out-vacuum wavefunctional. On $\Sigma_{\rm in}$, the restriction of the complex-conjugate in-mode is its boundary data, $u_k^{{\rm in}*}|_{\Sigma_{\rm in}}=q_k^{{\rm in}*}$. Since $K^{\rm \sigma}(\bm x,\bm y)=K^{\rm \sigma}(\bm y,\bm x)$ is symmetric, so is $\partial_{m^2}K^{\rm in}$. Hence the two terms in the symmetrized boundary two-point function contribute equally, canceling the factor $1/2$. Therefore
\begin{align}
\frac{i}{2}
\left\langle\varphi^{\rm in}\cdot\partial_{m^2}K^{\rm in}\cdot\varphi^{\rm in} \right \rangle
=\frac{i}{2}\sum_{kk'}(\alpha^{-1})_{kk'}
\int_{\Sigma_{\rm in}}\dd\Sigma_{\bm x}
\int_{\Sigma_{\rm in}}\dd\Sigma_{\bm y}\,u_{k'}^{{\rm out} }(\bm x)\partial_{m^2}K^{\rm in}(\bm x,\bm y) q_k^{{\rm in}*}(\bm y).
\end{align}
The kernel $K^{\rm in}$ satisfies the boundary equation obtained from the annihilation condition,
\begin{align}
\int_{\Sigma_{\rm in}}\dd\Sigma_{\bm y}\,
K^{\rm in}(\bm x,\bm y)q_k^{{\rm in} *}(\bm y)
=
p_k^{{\rm in} *}(\bm x).
\end{align}
Differentiating this equation with respect to $m^2$, while keeping the endpoint surface fixed, gives
\begin{align}
\int_{\Sigma_{\rm in}}\dd\Sigma_{\bm y}\,
\partial_{m^2}K^{\rm in}(\bm x,\bm y)q_k^{{\rm in} *}(\bm y)
=
\partial_{m^2}p_k^{{\rm in} *}(\bm x)-\int_{\Sigma_{\rm in}}\dd\Sigma_{\bm y}\,
K^{\rm in}(\bm x,\bm y)\partial_{m^2}q_k^{{\rm in} *}(\bm y).
\end{align}
The variation of the boundary data can be expanded in the complete boundary basis using the dual functions,
\begin{align}
\partial_{m^2}q_k^{{\rm in} *}(\bm y)= 
\sum_{j}\int_{\Sigma_{\rm in}}\dd \Sigma_{\bm z}q^{{\rm in}*}_j(\bm y)(q^{-1})^{{\rm in}*}_j(\bm z)\partial_{m^2}q_k^{{\rm in}*}(\bm z).
\end{align}
Substituting this expansion into the differentiated kernel equation rewrites the kernel variation as
\begin{align}
\int_{\Sigma_{\rm in}}\dd\Sigma_{\bm y}\,
\partial_{m^2}K^{\rm in}(\bm x,\bm y)q_k^{{\rm in} *}(\bm y)
=
\partial_{m^2}p_k^{{\rm in} *}(\bm x)-\sum_j p_j^{{\rm in}*}(\bm x)\int_{\Sigma_{\rm in}}\dd \Sigma_{\bm z}(q^{-1})^{{\rm in}*}_j(\bm z)\partial_{m^2}q_k^{{\rm in}*}(\bm z).
\end{align}
Inserting this identity into the kernel-variation contribution gives
\begin{align}
&\frac{i}{2}
\left\langle\varphi^{\rm in}\cdot\partial_{m^2}K^{\rm in}\cdot\varphi^{\rm in} \right \rangle \notag \\
=&\frac{i}{2}\sum_{kk'}(\alpha^{-1})_{kk'}
\int_{\Sigma_{\rm in}}\dd\Sigma_{\bm x}\, u_{k'}^{{\rm out} }(\bm x)
\left[ \partial_{m^2}p_k^{{\rm in} *}(\bm x)-\sum_j p_j^{{\rm in}*}(\bm x)\int_{\Sigma_{\rm in}}\dd \Sigma_{\bm z}(q^{-1})^{{\rm in}*}_j(\bm z)\partial_{m^2}q_k^{{\rm in}*}(\bm z)\right].
\end{align}
The remaining term involving \(p_j^{{\rm in}*}\) is reorganized by using the Wronskian relation $W_\Sigma[u_{k'}^{\rm out},u_k^{{\rm in}*}]=i\alpha_{k'k}$. Equivalently, on \(\Sigma_{\rm in}\) this relation implies
\begin{align}
\int_{\Sigma_{\rm in}}\dd \Sigma_{\bm x}u_{k'}^{\rm out}(\bm x)p_j^{{\rm in}*}(\bm x)=i\alpha_{k'j}+\int_{\Sigma_{\rm in}}\dd \Sigma_{\bm x} n_{\mu}\nabla^{\mu}u_{k'}^{\rm out}(\bm x) q^{{\rm in}*}_j(\bm x).
\end{align}
Using this identity, the Gaussian-kernel contribution from the in-vacuum wavefunctional becomes the sum of an endpoint Wronskian term and a boundary trace term:
\begin{align}
\frac{i}{2}
\left\langle\varphi^{\rm in}\cdot\partial_{m^2}K^{\rm in}\cdot\varphi^{\rm in} \right \rangle
=&\frac{i}{2}\sum_{kk'}(\alpha^{-1})_{kk'}W_{\Sigma_{\rm in}}\left[u_{k'}^{\rm out},\partial_{m^2}u_{k}^{{\rm in} *} \right]\notag\\
&+\frac{1}{2}\sum_k\int_{\Sigma_{\rm in}}\dd\Sigma_{\bm x} \left(q^{-1}\right)_k^{{\rm in}*}(\bm x)\partial_{m^2}q^{{\rm in}*}_k(\bm x).
\end{align}
The calculation for Gaussian-kernel contribution from the out-vacuum wavefunctional is parallel and gives
\begin{align}
-\frac{i}{2}\left \langle \varphi_{{\rm out}}\cdot \partial_{m^2}K^{{\rm out}*}\cdot\varphi^{\rm out} \right \rangle
=&\frac{i}{2}\sum_{kk'}\left( \alpha^{-1} \right)_{kk'}W_{\Sigma_{\rm out}}\left[\partial_{m^2}u_{k'}^{\rm out},u^{{\rm in}*}_k \right] \notag\\
&+\frac{1}{2}\sum_{k'}\int_{\Sigma_{\rm out}}\dd\Sigma_{\bm x}
\left(q^{-1}\right)_{k'}^{{\rm out}}(\bm x)\partial_{m^2}q^{{\rm out}}_{k'}(\bm x).
\end{align}

Adding the normalization contribution to the Gaussian-kernel contribution, the full mass variation of the in-vacuum wavefunctional is
\begin{align}
\left\langle
\partial_{m^2}\log\Psi_{\rm in}\right\rangle
=&
\frac{i}{2}\sum_{kk'}(\alpha^{-1})_{kk'}W_{\Sigma_{\rm in}}\left[u_{k'}^{\rm out},\partial_{m^2}u_{k}^{{\rm in} *} \right]\notag\\
&+\frac{1}{4}
\sum_k\int_{\Sigma_{\rm in}}\dd\Sigma_{\bm x} \left[ \left(q^{-1}\right)^{{\rm in} *}_{k}(\bm x)\partial_{m^2}q_k^{{\rm in} *}(\bm x)
-
 \left(q^{-1}\right)^{{\rm in}}_{k}(\bm x)\partial_{m^2}q_k^{{\rm in}}(\bm x)
\right] \notag \\
&+i\partial_{m^2}\theta^{\rm in},
\label{eq.Psi-in}
\end{align}
and the full mass variation of the out-vacuum wavefunctional is
\begin{align}
\left\langle
\partial_{m^2}\log\Psi^*_{\rm out}\right\rangle
=&
\frac{i}{2}\sum_{kk'}(\alpha^{-1})_{kk'}W_{\Sigma_{\rm out}}\left[\partial_{m^2}u_{k'}^{\rm out},u_{k}^{{\rm in} *} \right]\notag\\
&+\frac{1}{4}
\sum_{k'}\int_{\Sigma_{\rm out}}\dd\Sigma_{\bm x} \left[ \left(q^{-1}\right)^{{\rm out} }_{k'}(\bm x)\partial_{m^2}q_{k'}^{{\rm out} }(\bm x)
-
\left(q^{-1}\right)^{{\rm out}*}_{k'}(\bm x)\partial_{m^2}q_{k'}^{{\rm out}*}(\bm x)
\right] \notag \\
&-i\partial_{m^2}\theta^{\rm out}.
\label{eq.Psi-out}
\end{align}
Substituting \eqref{eq.Psi-in} and \eqref{eq.Psi-out} into \eqref{eq.W-Psi}, namely $\partial_{m^2}W_\Psi=-i\langle\partial_{m^2}\log\Psi_{\rm in}\rangle -i\langle\partial_{m^2}\log\Psi_{\rm out}^*\rangle$, gives
\begin{align}
\partial_{m^2}W_{\Psi}=
&\frac{1}{2}\sum_{kk'}(\alpha^{-1})_{kk'}W_{\Sigma_{\rm in}}\left[u_{k'}^{\rm out},\partial_{m^2}u_{k}^{{\rm in} *} \right]+\frac{1}{2}\sum_{kk'}(\alpha^{-1})_{kk'}W_{\Sigma_{\rm out}}\left[\partial_{m^2}u_{k'}^{\rm out},u_{k}^{{\rm in} *} \right] \notag\\
&+\frac{1}{2}\im \left[\sum_k\int_{\Sigma_{\rm in}}\dd\Sigma_{\bm x}  \left(q^{-1}\right)^{{\rm in} *}_{k}(\bm x)\partial_{m^2}q_k^{{\rm in} *}(\bm x) \right]\notag\\
&+\frac{1}{2}\im \left[\sum_{k'}\int_{\Sigma_{\rm out}}\dd\Sigma_{\bm x} \left(q^{-1}\right)^{{\rm out} }_{k'}(\bm x)\partial_{m^2}q_{k'}^{{\rm out} }(\bm x) \right] \notag\\
&+\partial_{m^2}\theta^{\rm in} -\partial_{m^2}\theta^{\rm out}.
\label{eq.W-Psi-Result}
\end{align}
In the last three lines of \eqref{eq.W-Psi-Result}, the quantities are real. Indeed, the terms written as $\im[\cdots]$ are real by definition, while $\theta^{\rm in}$ and $\theta^{\rm out}$ are arbitrary real phases. Therefore these terms contribute only to the real part of $W_\Psi$. The imaginary part of the endpoint contribution is controlled by the Wronskian terms in the first line.

\subsection{C. Boundary-Completed Effective Action and Vacuum Persistence}
The full in-out effective action contains both the bulk action and the endpoint vacuum wavefunctionals. Combining the bulk and endpoint wavefunctional contributions,
\begin{align}
\partial_{m^2}W= \partial_{m^2}W_{\rm G}+\partial_{m^2}W_{\Psi}.
\end{align}
Substituting \eqref{eq.WG} and \eqref{eq.W-Psi-Result}, the endpoint Wronskian terms cancel between the two contributions. One is then left with
\begin{align}
\partial_{m^2}W =\frac{i}{2}\sum_{k,k'}(\alpha^{-1})_{kk'}\partial_{m^2}\alpha_{k'k} +\partial_{m^2}\Theta.
\label{eq.MassVarW}
\end{align}
Here $\Theta$ is an arbitrary real phase, collecting the remaining real boundary contributions and the phase contributions from the normalization of the endpoint wavefunctionals in \eqref{eq.W-Psi-Result}
\begin{align}
\partial_{m^2}\Theta=&+\frac{1}{2}\im \left[\sum_k\int_{\Sigma_{\rm in}}\dd\Sigma_{\bm x}  \left(q^{-1}\right)^{{\rm in} *}_{k}(\bm x)\partial_{m^2}q_k^{{\rm in} *}(\bm x) \right]\notag\\
&+\frac{1}{2}\im \left[\sum_{k'}\int_{\Sigma_{\rm out}}\dd\Sigma_{\bm x} \left(q^{-1}\right)^{{\rm out} }_{k'}(\bm x)\partial_{m^2}q_{k'}^{{\rm out} }(\bm x) \right] \notag\\
&+\partial_{m^2}\theta^{\rm in} -\partial_{m^2}\theta^{\rm out}.
\end{align}
Its arbitrariness reflects the freedom to choose the phases of the normalized endpoint wavefunctionals. It shifts only the real part of the effective action. 
\begin{align}
W(m^2)=\frac{i}{2}\tr \log \alpha(m^2) +\Theta(m^2)+W_0,
\end{align}
where $W_0$ is independent of $m^2$ and is fixed by the choice of reference condition. Taking the large-mass limit as the reference condition, particle production is suppressed and $\alpha\alpha^\dagger\to I$. Hence $\operatorname{Im}W_0=0$. Since $\Theta$ is real, the imaginary part of the effective action reduces to
\begin{align}
\im W= \frac{1}{4}\tr \log \left[\alpha\alpha^{\dagger}\right].
\end{align}
Thus the imaginary part of effective action reduces to the Bogoliubov expression. The result can be checked directly at the differential level. Taking the mass derivative of the expression above gives
\begin{align}
\partial_{m^2}\im W
&=
\frac{1}{4}\tr\left[
(\alpha\alpha^\dagger)^{-1}
\partial_{m^2}(\alpha\alpha^\dagger)
\right]
\notag\\
&=
\frac{1}{4}\tr\left[
\alpha^{-1}\partial_{m^2}\alpha
+
(\alpha^\dagger)^{-1}\partial_{m^2}\alpha^\dagger
\right]
\notag\\
&=
\frac{1}{2}\re\tr\left[
\alpha^{-1}\partial_{m^2}\alpha
\right].
\end{align}
On the other hand, taking the imaginary part of \eqref{eq.MassVarW}
and using the reality of $\Theta$, gives
\begin{align}
\partial_{m^2}\im W=
\frac12\re\tr\left[
\alpha^{-1}\partial_{m^2}\alpha
\right].
\end{align}
The two expressions therefore agree identically.

\section{III. Illustrations of Boundary Completion}
\subsection{A. Time-Dependent Harmonic Oscillator: A Minimal Model}

In the main text, we formulated the boundary-completion mechanism for a free field on a general globally hyperbolic spacetime. The general treatment accommodates arbitrary Cauchy surfaces and mode mixing, but the associated functional and matrix structures may tend to obscure the underlying mechanism. Here we strip away these structures and consider a single time-dependent harmonic oscillator, which may also be viewed as a minimal analogue of the cosmological setting considered in Ref.~\cite{Akhmedov:2024qvi}. In this minimal setting, the Green's function integral reduces directly to endpoint Wronskians, allowing the Bogoliubov contribution and the additional endpoint terms to be identified in a straightforward manner. The latter are then canceled by the variation of the endpoint vacuum wavefunctions. We finally verify the completed result independently by evaluating the full Gaussian transition amplitude.

Consider the time-dependent harmonic oscillator with action
\begin{align}
S[\phi]
=
\frac{1}{2}\int_{t_i}^{t_f}\dd t\,
\left[
\dot{\phi}^2
-
\left(m^2+\omega^2(t)\right)\phi^2
\right].
\label{eq:osc-action}
\end{align}
A dot denotes differentiation with respect to time, $\dot{\phi}\equiv\partial_t\phi$. Here $m^2$ is a constant parameter with respect to which the effective action will be varied, while $\omega^2(t)$ is a prescribed time-dependent function independent of $m^2$.
Let $u^{\rm in}(t)$ and $u^{\rm out}(t)$ be the normalized positive-frequency modes defining the initial and final vacua. They satisfy
$\left[\partial_t^2+m^2+\omega^2(t)\right]u^\sigma=0$
and $W_t[u^\sigma,u^{\sigma *}]=i$, where
$W_t[f,g]\equiv f(t)\dot g(t)-\dot f(t) g(t)$. With
$u^{\rm out}=\alpha u^{\rm in}+\beta u^{{\rm in}*}$, one has
$W_t[u^{\rm out},u^{{\rm in}*}]=i\alpha$.

For a single oscillator, the coincident in-out Green's function is
$G_F(t,t)=\alpha^{-1}u^{\rm out}(t)u^{{\rm in}*}(t)$, so that
$\partial_{m^2}W_{\rm G}
=-\frac{1}{2\alpha}\int_{t_i}^{t_f}\dd t\,
u^{\rm out}u^{{\rm in}*}$.
The origin of the endpoint terms is then particularly transparent. Differentiating the equation for $u^{{\rm in}*}$ with respect to $m^2$ and using the equation for $u^{\rm out}$ gives
\begin{align}
\frac{\dd}{\dd t}
W_t\left[
u^{\rm out},
\partial_{m^2}u^{{\rm in}*}
\right]
=
-u^{\rm out}u^{{\rm in}*}.
\label{eq:osc-GL}
\end{align}
Thus the bulk Green's-function integral is reduced directly to the difference of this Wronskian at $t_f$ and $t_i$. Differentiating $W[u^{\rm out},u^{{\rm in}*}]=i\alpha$ then separates the Bogoliubov variation from the two endpoint pieces:
\begin{align}
\partial_{m^2}W_{\rm G}
=
\frac{i}{2\alpha}\partial_{m^2}\alpha
-
\frac{1}{2\alpha}
W_{t_i}\left[
u^{\rm out},
\partial_{m^2}u^{{\rm in}*}
\right]
-
\frac{1}{2\alpha}
W_{t_f}\left[
\partial_{m^2}u^{\rm out},
u^{{\rm in}*}
\right].
\label{eq:osc-WG-decomp}
\end{align}
This is the single-mode version of the boundary obstruction obtained in the general calculation.

The endpoint completion is equally simple. The normalized initial vacuum wavefunction is
\begin{align}
\Psi_{\rm in}(\phi_i)
=
\mathcal N_{\rm in}
\exp\left[
\frac{i}{2}
\frac{\dot u^{{\rm in}*}(t_i)}
     {u^{{\rm in}*}(t_i)}
\phi_i^2
\right],
\qquad
\mathcal N_{\rm in}
=
\left[2\pi|u^{\rm in}(t_i)|^2\right]^{-1/4}
e^{i\theta_{\rm in}} .
\label{eq:osc-in-wavefunction}
\end{align}
Its contribution to the mass variation separates into the Gaussian-kernel and normalization parts,
$-i\langle\partial_{m^2}\log\Psi_{\rm in}\rangle
=
-i\partial_{m^2}\log\mathcal N_{\rm in}
+
\frac{1}{2}\langle\phi_i^2\rangle
\partial_{m^2}
\left(
\frac{\dot u^{{\rm in}*}(t_i)}
     {u^{{\rm in}*}(t_i)}
\right)$.

For the Gaussian-kernel part,
$\langle\phi_i^2\rangle
=
G_F(t_i,t_i)
=
\frac{1}{\alpha}
u^{\rm out}(t_i)u^{{\rm in}*}(t_i)$
and
$\partial_{m^2}
\left(
\frac{\dot u^{{\rm in}*}}
     {u^{{\rm in}*}}
\right)
=
\frac{
W[u^{{\rm in}*},\partial_{m^2}u^{{\rm in}*}]
}{
(u^{{\rm in}*})^2
}$.
Its relation to the endpoint Wronskian in Eq.~\eqref{eq:osc-WG-decomp} follows from
\begin{align}
W_{t_i}\left[
u^{\rm out},
\partial_{m^2}u^{{\rm in}*}
\right]
={}&
\frac{u^{\rm out}(t_i)}
     {u^{{\rm in}*}(t_i)}
W_{t_i}\left[
u^{{\rm in}*},
\partial_{m^2}u^{{\rm in}*}
\right]
+
i\alpha
\frac{\partial_{m^2}u^{{\rm in}*}(t_i)}
     {u^{{\rm in}*}(t_i)} .
\label{eq:osc-W-recombine}
\end{align}
The Gaussian-kernel contribution is therefore
$\frac{1}{2\alpha}
W_{t_i}
[u^{\rm out},\partial_{m^2}u^{{\rm in}*}]
-
\frac{i}{2}
\frac{\partial_{m^2}u^{{\rm in}*}(t_i)}
     {u^{{\rm in}*}(t_i)}$.
The second term combines with the normalization variation
$-i\partial_{m^2}\log\mathcal N_{\rm in}
=
\frac{i}{4}
\left[
\frac{\partial_{m^2}u^{\rm in}(t_i)}
     {u^{\rm in}(t_i)}
+
\frac{\partial_{m^2}u^{{\rm in}*}(t_i)}
     {u^{{\rm in}*}(t_i)}
\right]
+
\partial_{m^2}\theta_{\rm in}$.
Defining
$\partial_{m^2}\Theta_{\rm in}
=
\partial_{m^2}\theta_{\rm in}
+
\frac{i}{4}
\left[
\frac{\partial_{m^2}u^{\rm in}(t_i)}
     {u^{\rm in}(t_i)}
-
\frac{\partial_{m^2}u^{{\rm in}*}(t_i)}
     {u^{{\rm in}*}(t_i)}
\right]$,
the complete initial-state contribution becomes
\begin{align}
-i\left\langle
\partial_{m^2}\log\Psi_{\rm in}
\right\rangle
=
\frac{1}{2\alpha}
W_{t_i}\left[
u^{\rm out},
\partial_{m^2}u^{{\rm in}*}
\right]
+
\partial_{m^2}\Theta_{\rm in}.
\label{eq:osc-in-W}
\end{align}

The final endpoint is completely parallel. For
$\Psi_{\rm out}^{*}
=
\mathcal N_{\rm out}^{*}
\exp[
-\frac{i}{2}
\frac{\dot u^{\rm out}(t_f)}
     {u^{\rm out}(t_f)}
\phi_f^2]$,
one obtains
\begin{align}
-i\left\langle
\partial_{m^2}\log\Psi_{\rm out}^{*}
\right\rangle
=
\frac{1}{2\alpha}
W_{t_f}\left[
\partial_{m^2}u^{\rm out},
u^{{\rm in}*}
\right]
+
\partial_{m^2}\Theta_{\rm out},
\label{eq:osc-out-W}
\end{align}
with
$\partial_{m^2}\Theta_{\rm out}
=
-\partial_{m^2}\theta_{\rm out}
-
\frac{i}{4}
\left[
\frac{\partial_{m^2}u^{\rm out}(t_f)}
     {u^{\rm out}(t_f)}
-
\frac{\partial_{m^2}u^{{\rm out}*}(t_f)}
     {u^{{\rm out}*}(t_f)}
\right]$.

Combining the bulk and endpoint contributions, the two endpoint Wronskians cancel term by term:
\begin{align}
\partial_{m^2}W
={}&
\underbrace{
\frac{i}{2\alpha}\partial_{m^2}\alpha
-
\frac{1}{2\alpha}
W_{t_i}\left[
u^{\rm out},
\partial_{m^2}u^{{\rm in}*}
\right]
-
\frac{1}{2\alpha}
W_{t_f}\left[
\partial_{m^2}u^{\rm out},
u^{{\rm in}*}
\right]}_{\partial_{m^2}W_{\rm G}}
\nonumber\\
&+
\underbrace{
\frac{1}{2\alpha}
W_{t_i}\left[
u^{\rm out},
\partial_{m^2}u^{{\rm in}*}
\right]
+
\frac{1}{2\alpha}
W_{t_f}\left[
\partial_{m^2}u^{\rm out},
u^{{\rm in}*}
\right]
+
\partial_{m^2}\Theta}_{\partial_{m^2}W_{\Psi}}
\nonumber\\
={}&
\frac{i}{2\alpha}\partial_{m^2}\alpha
+
\partial_{m^2}\Theta ,
\label{eq:osc-completion}
\end{align}
where
$\partial_{m^2}\Theta
=
\partial_{m^2}\Theta_{\rm in}
+
\partial_{m^2}\Theta_{\rm out}$
is real. The oscillator thus makes the boundary-completion mechanism transparent: the Green–Lagrange identity converts the bulk integral into the Bogoliubov contribution together with additional endpoint Wronskian terms, while the variation of the normalized endpoint vacuum states supplies the corresponding Wronskian terms with the opposite sign. Their cancellation leaves the Bogoliubov contribution, together with the real endpoint phase $\partial_{m^2}\Theta$.

The field-theory result can be viewed as the multimode generalization of this simplest example. In passing from the oscillator to field theory, the single-mode quantities are promoted to their corresponding mode-space objects, quotients are replaced by the appropriate inverse or dual structures, and the corresponding contractions and sums over all field modes are performed. Under these replacements, the entire oscillator calculation maps onto the general field-theory derivation of the effective action presented in the main text.

As an independent check, we can evaluate the full vacuum amplitude directly. Including both endpoint vacuum wavefunctions, the transition amplitude is
\begin{align}
Z
=
\langle {\rm out},t_f|{\rm in},t_i\rangle
=
\int \dd\phi_f\,\dd\phi_i\,
\Psi_{\rm out}^{*}(\phi_f)
K(\phi_f,t_f;\phi_i,t_i)
\Psi_{\rm in}(\phi_i).
\label{eq:osc-Z-gaussian}
\end{align}
Here $K(\phi_f,t_f;\phi_i,t_i)=\int_{\phi(t_i)=\phi_i}^{\phi(t_f)=\phi_f}\mathcal{D}\phi e^{iS[\phi]}$ is the propagator with fixed endpoint values. Since the theory is quadratic, this propagator is determined exactly by the classical solution. Using $\{u^{\rm in},u^{{\rm in}*}\}$ as a convenient basis and defining
$\Delta 
\equiv
u^{\rm in}(t_f)u^{{\rm in}*}(t_i)
-
u^{{\rm in}*}(t_f)u^{\rm in}(t_i)$,
the solution satisfying
$\phi_{\rm cl}(t_i)=\phi_i$
and
$\phi_{\rm cl}(t_f)=\phi_f$
can be constructed as
$\phi_{\rm cl}(t)
=
\phi_i
\frac{
u^{\rm in}(t_f)u^{{\rm in}*}(t)
-
u^{{\rm in}*}(t_f)u^{\rm in}(t)
}{\Delta}
+
\phi_f
\frac{
u^{\rm in}(t)u^{{\rm in}*}(t_i)
-
u^{{\rm in}*}(t)u^{\rm in}(t_i)
}{\Delta}$.
On shell,
$S_{\rm cl}
=
\frac{1}{2}
[\phi_{\rm cl}\dot{\phi}_{\rm cl}]_{t_i}^{t_f}$,
and the exact fixed-endpoint propagator is therefore
\begin{align}
K(\phi_f,t_f;\phi_i,t_i)
=
\frac{1}{\sqrt{-2\pi\Delta}}
\exp\left\{
\frac{i}{2\Delta}
\left[
\left(
\dot u^{\rm in}(t_f)u^{{\rm in}*}(t_i)
-
\dot u^{{\rm in}*}(t_f)u^{\rm in}(t_i)
\right)\phi_f^2
\right.\right.
\nonumber\\
\left.\left.
-
\left(
u^{\rm in}(t_f)\dot u^{{\rm in}*}(t_i)
-
u^{{\rm in}*}(t_f)\dot u^{\rm in}(t_i)
\right)\phi_i^2
+
2i\phi_f\phi_i
\right]
\right\}.
\label{eq:osc-propagator}
\end{align}
The square-root branch is fixed by the short-time prescription and continued continuously to $t_f$.

Substituting Eq.~\eqref{eq:osc-propagator} together with the endpoint vacuum wavefunctions into Eq.~\eqref{eq:osc-Z-gaussian}, the remaining integral over $\phi_f$ and $\phi_i$ is Gaussian. Writing its exponent as $\frac{i}{2}(\phi_f,\phi_i)M(\phi_f,\phi_i)^T$, one finds
\begin{align}
M
=
\begin{pmatrix}
\displaystyle
\frac{
\dot u^{\rm in}(t_f)u^{{\rm in}*}(t_i)
-
\dot u^{{\rm in}*}(t_f)u^{\rm in}(t_i)
}{\Delta}
-
\frac{\dot u^{\rm out}(t_f)}
     {u^{\rm out}(t_f)}
&
\displaystyle
\frac{i}{\Delta}
\\[3mm]
\displaystyle
\frac{i}{\Delta}
&
\displaystyle
-
\frac{
u^{\rm in}(t_f)\dot u^{{\rm in}*}(t_i)
-
u^{{\rm in}*}(t_f)\dot u^{\rm in}(t_i)
}{\Delta}
+
\frac{\dot u^{{\rm in}*}(t_i)}
     {u^{{\rm in}*}(t_i)}
\end{pmatrix}.
\label{eq:osc-M}
\end{align}
Although the individual entries depend on the endpoint values of the mode functions, their combination in the determinant simplifies considerably. Using
$u^{\rm out}
=
\alpha u^{\rm in}
+
\beta u^{{\rm in}*}$,
$W_t[u^{\rm out},u^{{\rm in}*}]=i\alpha$, and
$W_t[u^{\rm out},u^{\rm in}]=-i\beta$, one obtains
\begin{align}
\Delta\det M
=
\frac{
W_t[u^{\rm out},u^{{\rm in}*}]
}{
i\,u^{\rm out}(t_f)u^{{\rm in}*}(t_i)
}
=
\frac{\alpha}{
u^{\rm out}(t_f)u^{{\rm in}*}(t_i)
}.
\label{eq:osc-detM}
\end{align}

The remaining Gaussian integral is
$\int\dd\phi_f\,\dd\phi_i\,
\exp[\frac{i}{2}(\phi_f,\phi_i)M(\phi_f,\phi_i)^T]
=
2\pi i/\sqrt{\det M}$, where the square-root branch follows the same continuation prescription as the propagator.
Combining this result with Eq.~\eqref{eq:osc-detM} and the endpoint normalization factors gives
\begin{align}
Z
=
\mathcal N_{\rm out}^{*}\mathcal N_{\rm in}
\frac{1}{\sqrt{-2\pi\Delta}}
\frac{2\pi i}{\sqrt{\det M}}
=
\alpha^{-\frac{1}{2}}e^{i\Theta}.
\label{eq:osc-Z-product}
\end{align}
Thus the direct Gaussian evaluation makes explicit that the endpoint vacuum wavefunctions are an essential part of the vacuum amplitude. In particular, both their Gaussian kernels and their normalization factors are required: the former implement the in- and out-vacuum conditions at the two endpoints, while the latter cancel the moduli of the corresponding endpoint mode-function factors arising from the Gaussian determinant, with the remaining phase absorbed into $\Theta$. Omitting either contribution would therefore give an incomplete vacuum amplitude. Equation~\eqref{eq:osc-Z-product} immediately gives
$\operatorname{Im}W
=
\frac{1}{2}\log|\alpha|
=
\frac{1}{4}\log|\alpha|^2$,
in agreement with Eq.~\eqref{eq:osc-completion}.

\subsection{B. Completing the de Sitter Calculation}

We now complete the de Sitter calculation of Ref.~\cite{Zhou:2025jwm} by including the endpoint vacuum contributions. Consider the massive scalar field in the Poincar\'e patch,
$ds^2=\eta^{-2}(-d\eta^2+d\bm{x}^2)$, with $\eta\in(-\infty,0)$. We regulate the calculation by restricting the time interval to $\eta\in(\eta_i,\eta_f)$, with $\eta_i\to-\infty$ and $\eta_f\to0^-$, and imposing a comoving momentum cutoff $|k|\leq\Lambda$, with $\Lambda\to+\infty$. The time and momentum regulators define the dimensionless trade-off parameters $\lambda_i=-\eta_i\Lambda$ and $\lambda_f=-\eta_f\Lambda$. In the asymptotic limit, $\lambda_i\to+\infty$, whereas $\lambda_f\in\mathbb{R}_+$ parametrizes the interplay between the late-time and momentum cutoffs and hence the regularization scheme. We also use $\nu=\sqrt{m^2-d^2/4}$ and $C_d=2\pi^{d/2}/[(2\pi)^d\Gamma(d/2)]$.

With these regulators, the mass variation of the Green's function result obtained in Ref.~\cite{Zhou:2025jwm} reads
\begin{align}
\partial_{m^2}\im W_{\rm G}(\Lambda,\lambda_f)
=
\frac{\pi V_c C_d}
{2d(e^{2\pi\nu}-1)}
\left[
\frac{1}{(-\eta_f)^d}
\int_0^{\lambda_f}\dd x x^{d-1}\im J_{i\nu}(x)^2
+
\Lambda^d
\int_{\lambda_f}^\infty \frac{\dd x}{x}\im J_{i\nu}(x)^2
\right],
\label{eq:dS-WG-integral}
\end{align}
where $J_{i\nu}(x)$ is the Bessel function of the first kind. The second integral has a useful total-derivative structure. Defining $F_\nu(x)\equiv x\,\im W_x[J_{i\nu}(x),\partial_\nu J_{i\nu}(x)]$, with $W_x[f,g]=f\partial_xg-g\partial_xf$, the Bessel equation gives
\begin{align}
\frac{\dd F_\nu(x)}{\dd x}
=
-\frac{2\nu}{x}\im J_{i\nu}(x)^2.
\label{eq:dS-Q-identity}
\end{align}
Since $F_\nu(x)\to 1$ as $x\to\infty$, this gives $\int_{\lambda_f}^\infty \frac{\dd x}{x} \im J_{i\nu}(x)^2=[F_\nu(\lambda_f)-1]/(2\nu)$. Using $(-\eta_f)^{-d}=\Lambda^d/\lambda_f^d$, we obtain
\begin{align}
\partial_{m^2}\im W_{\rm G}(\Lambda,\lambda_f)
=
\frac{\pi V_c C_d\Lambda^d}
{4d\nu(e^{2\pi\nu}-1)}
\left[
F_\nu(\lambda_f)-1
+
\frac{2\nu}{\lambda_f^d}
\int_0^{\lambda_f}\dd x x^{d-1}\im J_{i\nu}(x)^2
\right].
\label{eq:dS-WG-Q}
\end{align}

We next evaluate the endpoint terms. Spatial homogeneity makes the Bogoliubov transformation diagonal in momentum space,
$\alpha_{\bm{k}'\bm{k}}=\alpha_k(2\pi)^d\delta^{(d)}(\bm{k}-\bm{k}')$, with
$|\alpha_k|^2=(1-e^{-2\pi\nu})^{-1}$. Substituting the de Sitter in- and out-modes given in Ref.\cite{Akhmedov:2024qvi,Zhou:2025jwm} directly into the boundary terms $\mathcal B_{\rm in}$ and $\mathcal B_{\rm out}$, the momentum delta functions reduce the sums over $\bm{k}$ and $\bm{k}'$ to a single momentum integral, with the coincident delta function giving the comoving spatial volume $V_c$. Performing the angular integration, $\int \frac{\dd^dk}{(2\pi)^d}=C_d\int \dd k k^{d-1}$, gives
\begin{align}
\im\mathcal B_{\rm in}(\Lambda,\lambda_i)
=
\frac{\pi V_c C_d\Lambda^d}
{4d\nu(e^{2\pi\nu}-1)}
\left[
F_\nu(\lambda_i)-1
+
\frac{2\nu}{\lambda_i^d}
\int_0^{\lambda_i}\dd x x^{d-1}\im J_{i\nu}(x)^2
\right],
\label{eq:dS-Bin}
\end{align}
and
\begin{align}
\im\mathcal B_{\rm out}(\Lambda,\lambda_f)
=
-\frac{\pi V_c C_d\Lambda^d}
{4d\nu(e^{2\pi\nu}-1)}
\left[
F_\nu(\lambda_f)
+
\frac{2\nu}{\lambda_f^d}
\int_0^{\lambda_f}\dd x x^{d-1}\im J_{i\nu}(x)^2
\right].
\label{eq:dS-Bout}
\end{align}

For the initial surface, $\eta_i\to-\infty$ at fixed $\Lambda$ implies $\lambda_i\to+\infty$. The large-$x$ behavior $F_\nu(x)=1+O(x^{-2})$, together with the oscillatory asymptotics of $\im J_{i\nu}(x)^2$, gives
\begin{align}
\lim_{\lambda_i\to+\infty}
\im\mathcal B_{\rm in}(\Lambda,\lambda_i)=0.
\label{eq:dS-Bin-zero}
\end{align}
The final-state contribution in Eq.~\eqref{eq:dS-Bout} then cancels the full $\lambda_f$ dependence of Eq.~\eqref{eq:dS-WG-Q}. For arbitrary finite $\lambda_f$ and $\Lambda$, we find
\begin{align}
\partial_{m^2}\im W_{\rm G}(\Lambda,\lambda_f)
+
0
+
\im\mathcal B_{\rm out}(\Lambda,\lambda_f)
=
-\frac{\pi V_c C_d\Lambda^d}
{4d\nu(e^{2\pi\nu}-1)}
=
\partial_{m^2}\im W_{\rm B}(\Lambda).
\label{eq:dS-boundary-completion}
\end{align}
Thus the dependence on the trade-off parameter $\lambda_f$, which parametrizes the regularization schemes considered in Ref.~\cite{Zhou:2025jwm}, is canceled by the final-state contribution before the regulators are removed. The boundary-completed result is independent of this regulator correlation and reproduces the Bogoliubov result.

\end{document}